\documentclass[superscriptaddress,preprintnumbers,amsmath,amssymb,aps,pre]{revtex4}
\usepackage{graphicx}
\usepackage{dcolumn}
\usepackage{bm}
\usepackage{color}
\usepackage{listings}
\usepackage[toc,page]{appendix}
\newcommand {\equalda} {\ {\raise-.5ex\hbox{$\buildrel d \over =$}}\ }
\begin{document}
\title{Temporal correlations between the earthquake magnitudes before major mainshocks in Japan}

\author{Panayiotis A. Varotsos}
\email{pvaro@otenet.gr} 
\affiliation{Section of Solid State Physics, Department of Physics, School of Science, National
and Kapodistrian University of Athens, Panepistimiopolis, Zografos
157 84, Athens, Greece}
\affiliation{Solid Earth Physics Institute, Department of
Physics, School of Science, National and Kapodistrian University
of Athens, Panepistimiopolis, Zografos 157 84, Athens, Greece}

\author{Nicholas V. Sarlis}
\email{nsarlis@phys.uoa.gr} 
\affiliation{Section of Solid State Physics, Department of Physics, School of Science, National
and Kapodistrian University of Athens, Panepistimiopolis, Zografos
157 84, Athens, Greece}
\affiliation{Solid Earth Physics Institute, Department of
Physics, School of Science, National and Kapodistrian University
of Athens, Panepistimiopolis, Zografos 157 84, Athens, Greece}

\author{Efthimios S. Skordas}
\email{eskordas@phys.uoa.gr} 
\affiliation{Section of Solid State Physics, Department of Physics, School of Science, National
and Kapodistrian University of Athens, Panepistimiopolis, Zografos
157 84, Athens, Greece}
\affiliation{Solid Earth Physics Institute, Department of
Physics, School of Science, National and Kapodistrian University
of Athens, Panepistimiopolis, Zografos 157 84, Athens, Greece}

\begin{abstract}
A characteristic change of seismicity has been recently uncovered when the precursory
Seismic Electric Signals activities initiate before an earthquake occurrence. In particular,
the fluctuations of the order parameter of seismicity exhibit a simultaneous 
distinct minimum upon analyzing the seismic catalogue in a new time domain termed natural
time and employing a sliding natural time window comprising a number of events that
would occur in a few months. Here, we focus on the minima preceding all earthquakes 
of magnitude 8 (and 9) class that occurred in Japanese area from 1 January 
1984 to 11 March 2011 (the day of the M9 Tohoku earthquake). By applying 
Detrended Fluctuation Analysis to the earthquake magnitude time series, we find that each of these minima is preceded as 
well as followed by characteristic changes of temporal correlations between 
earthquake magnitudes. In particular, we identify the following three main features.
The minima are observed during periods when long range 
correlations have been developed, but they are preceded by a stage in which an evident anti-correlated 
behavior appears. After the minima, the long range correlations 
break down to an almost random behavior turning to anti-correlation.  The minima 
that precede M$\geq$7.8 earthquakes are distinguished from 
other minima which are either non-precursory or 
followed by smaller earthquakes. 
\end{abstract}
\maketitle
\section{Introduction}
\label{intro}
Seismic Electric Signals (SES) are low frequency ($\leq 1$Hz)
transient changes of the electric field of the Earth that have
been found \citep{VAR84A,VAR96B} to precede earthquakes (EQs).
Several such transient changes within a short time are termed SES
activity. A  model for the SES generation has been
proposed \citep{VARBOOK} (see also \citet{VAR93}) based on the
widely accepted concept that the stress gradually increases in the
future focal region of an EQ. It was postulated that when this stress reaches a {\em
critical} value, a cooperative orientation of the electric dipoles
(which anyhow exist in the focal area due to lattice defects
in the ionic constituents of the rocks) occurs, which leads to the
emission of a transient electric signal. 
All solids including metals, insulators and semiconductors contain intrinsic and extrinsic defects\citep{VAROTSOS1974927,KOS75,VAR997,VARALEX80p,VAR78,VAR80K133,VARALEX82B,VARALEX82,VAR08438}. 
The model is consistent
with the finding that the time series of the observed SES
activities (along with their associated magnetic field variations)
exhibit infinitely ranged temporal correlations
\citep{NAT03A,NAT03B,NAT08,NAT09V}, thus being in accord with the
conjecture of {\em critical} dynamics. 
{Other possible mechanisms for SES generation such as the recently developed 
finite fault rupture model with the electrokinetic effect\citep{REN12} and the piezoelectric 
effect\citep{VARO78} taking into account the fault dislocation theory\citep{HUA11A} have been proposed, see also Ch. 1 of \citet{SPRINGER}.} 
The observations of SES
activities in Greece \citep{VAR91,VAR93,SPRINGER} have shown that
their lead time is of the order of a few months. This agrees with
later observations in {Japan \citep{UYE00,UYE02,UYE09,ORI12}. 

EQs may be considered as (non-equilibrium) critical phenomena
since the observed EQ scaling laws \citep{TUR97} point to
the existence of phenomena closely associated with the proximity
of the system to a critical point \citep{HOL06}. An order parameter
for seismicity has been introduced \citep{NAT05C} in the frame of
the analysis in a new time domain termed natural time $\chi$ (see
below). This analysis has been found to reveal novel dynamical
features hidden in the time series of complex systems
\citep{SPRINGER}.

A unique change of the order parameter of seismicity approximately
at the time when SES activities initiate has been recently
uncovered \citep{TECTO12}. In particular, upon analyzing the
Japanese seismic catalogue in natural time, and employing a
sliding natural time window comprising the number of events that
would occur in a few months, the following was observed: The
fluctuations of the order parameter of seismicity exhibit a
clearly detectable minimum approximately at the time of the
initiation of the pronounced SES activity observed by \citet{UYE02,UYE09} 
almost two months before the onset of
the volcanic-seismic swarm activity in 2000 in the Izu Island
region, Japan. (This swarm was then characterized by Japan
Meteorological Agency (JMA) as being the largest EQ swarm
ever recorded \citet{JMA00}.) This reflects that presumably the
same physical cause led to both effects, i.e, the emission of the
SES activity and the change of the correlation properties between
the EQs. In addition, these two phenomena were found
\citep{TECTO12} to be also linked in space.

For the vast majority of major EQs in Japan, however, the
aforementioned almost simultaneous appearance of the minima of the
fluctuations of the order parameter of seismicity with the
initiation of SES activities, cannot be directly verified due to
the lack of geolectrical data. In view of this lack of data, an
investigation was made \citep{PNAS13} that was solely focused on
the question whether minima of the fluctuations of seismicity are
observed before all  EQs of magnitude 7.6 or larger that
occurred from 1 January 1984 to 11 March 2011 (the day of the M9
Tohoku EQ) in Japanese area. Actually such minima were identified
a few months before these EQs. It is the main scope of
this paper to investigate the temporal correlations between the
EQ magnitudes by paying attention to the time periods
during which the minima of the order parameter fluctuations of
seismicity have been observed before EQs of magnitude 8
(and 9) class {(cf. \citet{PNAS13} also studied the M7.6 Far-Off Sanriku EQ which 
however is not studied in detail here -but only shortly commented in the last paragraph of the Appendix- since it belongs to a smaller magnitude class, i.e., the 7-7.5 class, being a single asperity event\citep{YAM04})}.   Their epicenters are shown in Fig. \ref{fig1} (see
also Table \ref{tab1}). Along these lines, we employ here the
Detrended Fluctuation Analysis (DFA) \citep{PEN94} which has been
established as a standard method to investigate long range
correlations in non-stationary time series in diverse fields{
(e.g., \citet{PEN93,PEN94,PEN95B,ASH02,IVA09,IVA07B,TAL00,GOL02,TEL09,TEL06,TEL12}) including the study of geomagnetic data associated with the M9.0 Tohoku EQ \citep{RON12}.}
For example, a  recent study \citep{SHAO12} showed that DFA as
well as the Centered Detrended Moving Average technique remain
``The Methods of Choice'' in determining the Hurst index of time
series.  As we shall see, the  results of DFA obtained here in
conjunction with the aforementioned minima emerged from natural
time analysis lead to conclusions that are of key importance for
EQ prediction research. In particular, we find that each
of these precursory minima {of the fluctuations of
the order parameter of seismicity} is preceded as well as followed
by characteristic changes of temporal correlations between
EQ magnitudes{, thus complementing the results of \citet{PNAS13}}.

\section{The procedure followed in the analysis}\label{sec2}
In a time series comprising $N$ consecutive events, the natural
time of the $k$-th event of energy $Q_k$ is defined by $\chi_k =
k/N$ \citep{NAT01,NAT02,NAT02A}. We then study the evolution of the pair
$(\chi_k,p_k)$, where $p_k={Q_k}/{\sum_{n=1}^NQ_n}$ is the
normalized energy. This analysis, termed natural time analysis,
extracts from a given complex time series the maximum information
possible \citep{ABE05}. The approach of a dynamical system to a
critical point can be identified \citep{NAT01,SPRINGER,PNAS} by
means of the variance $\kappa_1$ of natural time $\chi$ weighted
for $p_k$, namely
\begin{equation}\label{kappa1}
\kappa_1=\sum_{k=1}^N p_k (\chi_k)^2- \left(\sum_{k=1}^N p_k
\chi_k \right)^2 \equiv \langle \chi^2 \rangle - \langle \chi
\rangle^2
\end{equation}
It has been argued\citep{NAT05C} (see also pp. 249-253 of
\citet{SPRINGER}) that the quantity $\kappa_1$ of seismicity can
serve as an order parameter. To compute the fluctuations of
$\kappa_1$ we apply the following procedure
\citep{SPRINGER,PNAS13}: First, take an excerpt comprising $W (\geq
100)$ successive EQs from the seismic catalogue. We call it
excerpt $W$. Second, since at least 6 EQs are needed for
calculating reliable $\kappa_1$ \citep{NAT05C}, we form a window of
length 6 (consisting of the 1st to the 6th EQ in the excerpt $W$)
and compute $\kappa_1$ for this window. We perform the same
calculation of successively sliding this window through the whole
excerpt $W$. Then, we iterate the same process for windows with
length 7, 8 and so on up to $W$. (Alternatively, one may use
\citep{SPRINGER,VAR11,TECTO12} windows with length 6, 7, 8 and so on up
to $l$, where $l$ is markedly smaller than $W$, e.g., $l \approx
40$.) We then calculate the average value $\mu(\kappa_1)$ and the
standard deviation $\sigma(\kappa_1)$ of the ensemble of
$\kappa_1$ thus obtained. The quantity $\beta_{W} \equiv
\sigma(\kappa_1) / \mu(\kappa_1)$ is defined\citep{NEWEPL} as the variability of
$\kappa_1$ for this excerpt of length $W$ 
and is assigned to the $(W+1)^{th}$ EQ in the catalogue, the
target EQ. (Hence, for the $\beta_W$ value of a target EQ only its
past EQs are used in the calculation.) The time evolution of the
$\beta$ value can then be pursued by sliding the excerpt $W$
through the EQ catalogue and the corresponding minimum value (for
at least $W$ values before and $W$ values after) is labelled
$\beta_{W,min}$.

In addition, for the purpose of the present study, for the target
EQ we apply the standard procedure \citep{PEN94,GOL00} of DFA to
the magnitude time series of the {\em preceding} 300 EQs, which is
on the average the number of events that occurred  in the past few
months (see also below). Hence, for the target EQ we deduce a DFA
exponent, hereafter labelled $\alpha$ {(cf. $\alpha= $0.5
means random, $\alpha$ greater than 0.5  long range correlation, and $\alpha$ less than 0.5  anti-correlation.)}. By the same token the time
evolution of the $\alpha$ value can be pursued by sliding the
natural time window of length 300 through the EQ catalogue. The
minimum values $\alpha_{min}$ of the $\alpha$ exponent observed
(roughly three months) before (bef) and after (aft) the
identification of $\beta_{W,min}$ are designated by
$\alpha_{min,bef}$ and $\alpha_{min,aft}$, respectively (cf. when
a major  EQ takes place,  $\alpha_{min,aft}$ is the minimum $\alpha$
value after $\beta_{W,min}$ up to this EQ occurrence). {In particular, the $\alpha_{min,bef}$ and $\alpha_{min,aft}$ values (given in
detail in Tables \ref{tab2} to \ref{tab5}) were determined by investigating the minimum of the $\alpha$ exponent up to 
three and half months (105 days) before and after $\beta_{250,min}$, respectively.}

\section{The data analyzed} \label{sec3}  

The JMA seismic catalogue was used.
We considered all the EQs in the period from 1984 until the Tohoku
EQ occurrence on 11 March 2011, within the area $25^o - 46^o$N,
$125^o - 148^o$E shown by the black rectangle in Fig. 1. The
eastern edge of this area has been extended by $2^o$ to the E
compared to the area $25^o - 46^0$N, $125^o - 146^o$E (yellow
rectangle in Fig. 1) studied by \citet{TECTO12} for two
reasons: First, when plotting in Fig. 1 the links along with the
corresponding nodes recently identified by a network approach
developed by \citet{TEN12}, we see that the
nodes in the uppermost right part are now surrounded by the black
rectangle but not by the yellow one (cf. a node represents a
spatial location while a link between two nodes represents similar
seismic activity patterns in the two different locations
\citep{TEN12} ). Second, the epicenter of {the}
major EQ of magnitude 8.2 that occurred on 4 October 1994 lies
inside the former rectangle, but not in the latter (Table \ref{tab1}).

The energy of EQs was obtained from the magnitude M$_{JMA}$ reported by JMA after converting \citep{TAN04} to
the moment magnitude M$_w$\citep{KAN78}. Setting a threshold
M$_{JMA}$ = 3.5 to assure data completeness, we are left with
47,204 EQs and 41,277 EQs in the concerned period of about 326
months in the larger (black rectangle) and smaller (yellow
rectangle) area, respectively. Thus, we have on the average $\sim
145$ and $\sim 125$ EQs per month for the larger and smaller area,
respectively. In what follows, for the sake of
{brevity } in the
calculation of $\beta_W$ values for both areas,  we shall use the
values $W=200$ and $W=300$ (as in \citet{PNAS13}), which would cover a period of around a
few months before each target EQ. In addition for the sake of  comparison 
between the two areas, we will
also investigate the case of $W=250$ since this value in the larger area roughly corresponds
 to the case $W=200$ in the smaller area.

\section{Results}
Figure \ref{fig2} provides an overview of the values computed in
this study. In particular, the following quantities are plotted
versus the conventional time during the 27 year period from 1
January 1984 until the Tohoku EQ occurrence on 11 March 2011: In
Fig. \ref{fig2}{A} the DFA exponent $\alpha$ is depicted with
red line for the larger area and with green line  for the
smaller. In Fig. \ref{fig2}{B}, we show the quantities $\beta_{200}$
and $\beta_{300}$ (in red and blue, respectively) for the smaller
area. Finally, in Fig.
\ref{fig2}{C}, we show $\beta_{200}$, $\beta_{250}$ and
$\beta_{300}$ (in red, green and blue, respectively) for the
larger area. 

A first inspection of the $\alpha$ values in Fig. \ref{fig2}{A} 
shows that in view of their strong fluctuations it is very
difficult to identify their correlations with EQs. A closer
inspection, however, reveals the following striking point: The
deeper minima of the $\alpha$ values (when considering the $\alpha$ values 
in {\em both} areas) are
observed in the periods marked with grey shade which are very
close to the occurrence of the stronger EQs  in Japan
during the last decade. These two EQs are (Table \ref{tab1}) the M9 Tohoku EQ on 11
March 2011 with an epicenter at $38.10^o$N $142.86^o$E and the M8
Off-Tokachi EQ on 26 September 2003 with an epicenter at
$41.78^o$N $144.08^o$E . This instigated a more detailed
investigation of the $\alpha$ values close to these two major EQs
for which unfortunately precursory geoelectrical data are
lacking (for the case of the M9 Tohoku EQ only geomagnetic data are available, see below). 
Thus, before presenting these two investigations and in order to better understand 
the results obtained, we first describe below a similar investigation for the case of 
the volcanic-seismic swarm activity in 2000 in the Izu Island region, Japan, 
in which as mentioned both datasets, i.e., SES activities and seismicity, are available.

\subsection{The case of the volcanic-seismic swarm in 2000 in the Izu Island region}

Figure \ref{fig3} is an excerpt
of Fig. \ref{fig2} in expanded time scale during the six month
period from 1 January 2000 until 1 July 2000, which is the date of  occurrence of an
M6.5 EQ close to Niijima Island (yellow square in
Fig.\ref{fig1}). This EQ was preceded by an SES activity initiated
on 26 April 2000 at a measuring station located at this island \citep{UYE02,UYE09}.

An inspection of Figs. \ref{fig3}{A} to \ref{fig3}{C}
reveals the following three main features referring to the periods
before, during, and after  the observation of the
precursory $\beta$ minimum:

{Stage A} marked in cyan: Putting the details
aside, we observe in Fig. \ref{fig3}{A} that around 12
February 2000 the DFA exponent in both areas went down to a value
markedly smaller than 0.5, i.e., $\alpha \approx 0.41$, in the
smaller area and $\alpha \approx 0.43$ in the larger area. These
$\alpha_{min,bef}$ values indicate {anticorrelated} behavior in
the magnitude time series.

{Stage B} marked in yellow: Since the last days
of March until the first days of June 2000, the exponent $\alpha$
becomes markedly larger than 0.5, i.e., around $\alpha \approx
0.57$, pointing to the development of long range temporal
correlations. In Figs. \ref{fig3}{B} and \ref{fig3}{C}, we
then observe that after the last days of March the variability
$\beta$ exhibits a gradual decrease and a minimum $\beta_{W,min}$
appears on a date around the 
date  of the initiation of the SES activity. 
In particular, in  Fig. \ref{fig3}{C}, the relevant curve (green) for 
$\beta_{250}$ in the larger area minimizes on 25 April 2000 
which is approximately the date of the initiation of the SES activity, 
reported by \citet{UYE02,UYE09}, lying also very close to the date
(21 April) at which in the smaller area the $\beta_{200}$
curve (red) in Fig. \ref{fig3}B minimizes.
Thus, in short the minimum $\beta_{W,min}$, appears when $\alpha
> 0.5$ and hence when long range correlations (corr) have been developed 
in the EQ magnitude time series.
The corresponding $\alpha$ values during the observation of the
minima $\beta_{250,min}$ will be hereafter designated
$\alpha_{corr}$. Hence, in this case $\alpha_{corr} \approx 0.57$.

{Stage C} marked in brown: Approximately on 10
June 2000, Fig. \ref{fig3}{\em A} shows that the DFA exponent
decreases to a value around 0.5. This means that the previously
established long range temporal correlations between EQ
magnitudes break down to an almost random behavior. The value
$\alpha \approx 0.5$ remains almost constant until the third
week of June and shortly after the aforementioned M6.5 EQ 
 on 1 July 2000 occurred.

\subsection{The M8 Off Tokachi EQ on 26 September 2003}

Figure \ref{fig4} is an excerpt of Fig. \ref{fig2} in expanded
time scale which shows clearly what happened during an almost 6
month period before the occurrence of this major EQ. Using the
same symbols as in the description of the previous case, the
following three main features emerge from Figs. \ref{fig4}{A}
to \ref{fig4}{C}:

(A) An anti-correlated behavior is evident in the cyan 
region in Fig. \ref{fig4}A lasting from 1 April until 
the first days of May 2003, since the corresponding  $\alpha$ values 
resulting from both areas scatter around $\alpha=$0.42 and 0.44 for 
the larger and the smaller area, respectively.

(B) In the yellow region, i.e., from around the last days of May until
almost the beginning of the last week of July 2003, long range correlations
have been developed since the $\alpha$ values in Fig. \ref{fig4}A markedly 
exceed 0.5. The $\beta$ minima appear 
during this period (Tables \ref{tab2} and \ref{tab4}), as can be seen in Figs.\ref{fig4}C and \ref{fig4}B,
and the corresponding $\alpha$ values are $\alpha_{corr}\approx 0.6$.

(C) A breakdown of long range correlations starts 
around 1 September 2003, see the beginning of 
the brown region in Fig.\ref{fig4}A where the $\alpha$ values decrease to $\alpha \approx$0.5,
i.e., close to random behavior, and 
subsequently go down to around $\alpha \approx$0.45 
(while finally  $\alpha$ reaches the value $\alpha_{min,aft}=$0.384 and 0.434 for 
the larger and the smaller area, respectively, see Tables \ref{tab2} and \ref{tab4}) 
indicating anti-correlation. 
The M8 EQ occurred three weeks later, i.e., on 26 September 2003, 
and after its occurrence the $\alpha$ value decreases to
an unusual low $\alpha$ value (0.33 and 0.35 in the larger and the smaller area, respectively),
which corresponds to the one -out of the two- deeper $\alpha$ minima mentioned 
above in the periods marked with grey shade in Fig. \ref{fig2}A.

\subsection{The M9 Tohoku EQ on 11 March 2011}

Figure \ref{fig5} is an almost four month excerpt of Fig. \ref{fig2}
in which one can visualize what happened before this
{super giant} EQ. By the same token as in the
previous cases, the following three main features are recognized:

{(A)} The cyan region in Fig. \ref{fig5}
corresponds to an anticorrelated behavior since in Fig.
\ref{fig5}{A} the $\alpha$ values in both areas become markedly smaller than 0.5 
after around 16 December 2010, including an
evident minimum $\alpha_{min,bef}$ on 22 December 2010.
This is one out of the two deeper $\alpha$ minima mentioned in
Fig. \ref{fig2}{\em A}.

{(B)} In the yellow region, from about 23
December 2010 until {around 8} January 2011, the
$\alpha$ values depicted in Fig. \ref{fig5}{A} indicate the
establishment of long range correlations since $\alpha > 0.5$. In
this region, and in particular during the last week of December
2010, the $\beta$ values in Figs. \ref{fig5}{B} and
\ref{fig5}{C} show that an evident decrease starts leading to
a deep $\beta$ minimum around 5 January 2011. This is the deepest
$\beta_{W,min}$ observed \citep{PNAS13} since the beginning of our investigation on
1 January 1984, as
can be seen in the rightmost side of Figs. \ref{fig2}{B} and
\ref{fig2}{C}.
Remarkably, the anomalous magnetic field variations
\citep{XU013} 
(which accompany anomalous electric field variations, i.e., SES activities \citep{VAR03})
initiated almost on the same date, i.e., 4 January 2011.

{(C)} In the brown region lasting from about 13 January to 10 February 2011, 
the behavior turns to an anti-correlated one, 
which is very close to random, as evidenced from Fig.
\ref{fig5}{A} in which the $\alpha$ values are $\alpha
\lesssim 0.5$. The M9 EQ occurred {almost
four} weeks after this period, i.e., on 11 March 2011.

The following important comment referring to the two deeper
minima of the $\alpha$ values in Fig. \ref{fig2}{A} is now in
order: Here, the unusually low $\alpha_{min,bef}$ on 22 December
2010 (Fig. \ref{fig5}{A}) has been shortly followed by the
deep{est} $\beta$ minimum on 5 January 2011 (Figs.
\ref{fig5}{B} and \ref{fig5}{C}). This is of precursory
nature. To the contrary, the unusually low $\alpha$ minimum value
on 26 September 2003 discussed in the previous case -which has not
been shortly followed by a deep $\beta$ minimum (see Fig.
\ref{fig2}{A})- is not precursory {having been influenced by the preceding M8 Off Tokachi  EQ}. In other words, upon the
observation of an unusually low $\alpha$ value, we cannot decide
whether it is of precursory nature but we have to combine this
observation with the results of natural time analysis and
investigate whether this $\alpha_{min}$ value is shortly followed
by a deep $\beta_{W,min}$ value. {Hence, it is of
key} importance {to examine in each case}
 whether the sequence of the
aforementioned three main features {A},
{B}, {C} has appeared or
not.

\section{Discussion and conclusions}\label{Sec4}

{DFA has been employed long ago for the study of seimic time series in various regions, e.g. see \citet{TEL03} for the Italian territory.  
DFA studies of the long-term seismicity in Northern and Southern California  were initially focused on the 
regimes of stationary seismic activity }
and found that long range correlations exist
\citep{LEN08} between EQ magnitudes with $\alpha = 0.6$.
Similar DFA studies of long-term seismicity were later\citep{NEWEPL,NEWTSA} extended 
also to the seismic data of Japan and the results strengthened
the existence of long range temporal correlations. In particular, it was 
found\citep{NEWEPL,NEWTSA} that the DFA exponent
is around 0.6 for short scales but $\alpha=$0.8-0.9 for longer scales 
(the cross-over being around 200 EQs). In addition, the nonextensive statistical 
mechanics\citep{ABE01,TSA09}, pionered by \citet{TSA88}, has been 
employed\citep{NEWTSA} in order to investigate whether it can reproduce the observed seismic 
data fluctuations. In this framework, on the basis of which it has been 
shown \citep{LIV09} that kappa distributions arise, a generalization of the 
Guternbeng-Richter (G-R) law for seicmicity has been offered (for details and 
relevant references see Section 6.5 of \citet{SPRINGER} {as well as \citet{TEL12A}}) and the investigation led
to the following conclusions\citep{SPRINGER,NEWTSA}: The results of the natural time analysis of synthetic 
seismic data obtained from either the conventional G-R law or its nonextensive generalization,
deviate markedly from those of the real seismic data. On the other hand,
if temporal correlations between EQ magnitudes, with different $\alpha$ values (i.e., 
$\alpha\approx 0.6$ and $\alpha \approx$0.8-0.9 for short and long scales, respectively),
will be inserted to the synthetic seismic data, the results of natural time analysis agree 
well with those obtained from the real seismic data. In other words, the parameter $q$ 
of nonextensive statistical mechanics cannot capture the whole effect of long range 
temporal correlations between the magnitudes of successive EQs. On the other hand,
the nonextensive statistical mechanics, when combined with natural time analysis 
(which focuses on the sequential order of the events that appear in nature) does enable 
a satisfactory description of the fluctuations of the real data of long-term seismicity.

In the present paper, we study the dynamic evolution of seismicity and 
pay attention to the regimes before major EQs by combining the results of  DFA of
EQ magnitude time series  with natural
time analysis since the latter has revealed that a minimum
$\beta_{W,min}$ in the fluctuations of the order parameter of
seismicity is observed before major EQs in California \citep{EPL12}
and Japan \citep{PNAS13,TECTO12} (cf. The nonextensive statistical mechanics cannot 
serve for the purpose of the present study, i.e., follow the dynamic evolution 
of seismicity). This combination has been applied
in the previous Section to three characteristic cases in Japan,
i.e., the volcanic-seismic swarm activity in the Izu Island
region in 2000, the M8 off Tokachi EQ in 2003 and the M9 Tohoku EQ
in 2011. The following three main features have been found in all
three cases:

{Stage A}(before the $\beta_{W,min}$):
Clear anti-correlated behavior, $\alpha < 0.5$

{Stage B}: Establishment of long range
correlations, $\alpha > 0.5$ during which a minimum
$\beta_{W,min}$ appears (approximately at the date of the
initiation of the SES activity {as found in the
Izu case as well as in the M9 Tohoku case}).

{Stage C} (after the $\beta_{W,min}$): Breakdown
of long range correlations with emergence of an almost random 
behavior turning to anti-correlation, $\alpha \lesssim 0.5$.  
A few weeks after this breakdown the major EQ occurs. This is strikingly reminiscent 
of the findings in other complex time series: In the case of electrocardiograms, 
for example, the long-range temporal correlations that characterize 
the healthy heart rate variability break down for individuals of high
risk of sudden cardiac death and often 
accompanied by emergence of uncorrelated randomness
\citep{SPRINGER,GOL02,NAT07}.

The same features have been found to hold before all the other
${\rm M}_{JMA} \geq 7.8$  EQs in Japan  during the period 1 January
to the Tohoku EQ occurrence, i.e., the Southwest-Off Hokkaido M7.8
EQ in 1993 (Fig.\ref{figA1}) the East-Off Hokkaido M8.2 EQ in 1994
(Fig. \ref{figA2}),  and the Near Chichi-jima M7.8 EQ in 2010 (Fig. \ref{figA3}). {As for 
the observed pattern in $\alpha$, i.e., anti-correlated, correlated and random, it might be 
related to the tectonics and geodynamics, but a precise physical justification of its origin is not yet clear.}

The $\beta$ minima that are precursory to  ${\rm M}_{JMA} \geq
7.8$ EQs can be distinguished from other $\beta$ minima that are either non-precursory
or may be followed by EQs of smaller magnitude through the following procedure
 (for details see Appendix and
Tables \ref{tab2} to \ref{tab5}). We make separate studies 
for the two rectangular areas shown in Fig. \ref{fig1}, i.e., 
by analyzing the time series of EQs occurring in each area, 
first in the larger area and secondly in the smaller. 
In the study of each area, we do the following: We first 
identify the $\beta$ minima that appear 
a few months before all ${\rm
M}_{JMA} \geq 7.8$ EQs 
and determine their $\beta_{300,min}/\beta_{200,min}$ values. 
These values lie in a certain narrow range close to unity. 
Among the remaining minima, we choose those which are equally
deep or deeper than the shallowest one of the $\beta_{200,min}$ values
that preceded the ${\rm
M}_{JMA} \geq 7.8$ EQs 
and in addition they have  $\beta_{300,min}/\beta_{200,min}$ values 
lying in the range determined above (see Appendix).
In order for any of these minima to be precursory to ${\rm
M}_{JMA} \geq 7.8$ EQs, beyond the fact that 
they should exhibit the three main features, A, B, C, mentioned above, 
they should also have the following property: They should 
appear practically on the same dates (differing by no more than 10 days or so) 
in the investigations of both areas. 
The application of the aforementioned procedure reveals (see Appendix) 
that only the $\beta$  minima appearing a 
few months before the five ${\rm
M}_{JMA} \geq 7.8$  EQs exhibit all the aforementioned properties. 
Remarkably, this procedure (see Appendix) could have 
been applied before the occurrence of the M9 Tohoku EQ, 
after the identification of the deepest $\beta$ minimum 
observed on January 2011, leading to the conclusion that an
${\rm
M}_{JMA} \geq 7.8$  EQ was going to occur in a few months.

Let us summarize:  Here, by employing the DFA of the EQ
magnitude time series we show that the minimum $\beta_{W,min}$ of
the fluctuations of the order parameter of seismicity a few months before an 
${\rm M}_{JMA} \geq 7.8$  EQ is observed
when long range correlations prevail ($\alpha > 0.5$). In
addition, these $\beta_{W,min}$ is preceded by a stage in which DFA
reveals clear anti-correlated behavior ($\alpha < 0.5$) as well as
it is followed by another stage in which long range correlations
break down to an almost random behavior  turning to anti-correlation
($\alpha \lesssim 0.5$).
 On the basis of these main features we suggest a procedure which 
 distinguishes the 
 $\beta$ minima that precede  EQs of magnitude exceeding a certain threshold 
 (i.e., ${\rm
M}_{JMA} \geq 7.8$) from other $\beta$ minima which are either
non-precursory or may be followed by EQs of smaller magnitude.

\appendix
\section{Distinction of the $\beta$ minima that precede  EQs of
magnitude ${\rm M}_{JMA} \geq 7.8$ from other  minima which are
either non-precursory or followed by EQs of smaller
magnitude}

Recall that in order to classify a $\beta_{W,min}$
value, it should be a minimum for at least $W$ values before and
$W$ values after. Further, to assure that $\beta_{200,min}$,
$\beta_{250,min}$ and $\beta_{300,min}$ are precursory of the same
mainshock and hence belong to the same critical process, almost
all (in practice above 90$\%$ of) the events which led to
$\beta_{200,min}$ should participate in the calculation of
$\beta_{250,min}$ and $\beta_{300,min}$.

To distinguish the $\beta_{W,min}$ that are precursory to 
EQs of magnitude ${\rm M}_{JMA} \geq 7.8$ from other
minima which are either non-precursory or may be followed by EQs
of smaller magnitude, we make separate studies for the larger and
the smaller area and the results obtained should be necessarily
checked for their self-consistency. For example, a major EQ whose
epicenter lies in both areas should be preceded by $\beta_{W,min}$
identified in the separate studies of these two areas
approximately (in view of their difference in seismic rates) on
the same date. In particular,  we work as follows:

Let us assume that we start the investigation from the larger area
where five  EQs of magnitude 7.8 or larger occurred from 1 January
1984 until the M9 Tohoku EQ in 2011 (Table \ref{tab1}). We first
identify the $\beta$ minima that appear a few months before all
these EQs and determine their $\beta_{300,min} / \beta_{200,min}$
values (see Table \ref{tab2}). These values are found to lie in a
narrow range close to unity \citep{PNAS13}, i.e., in the range
0.92-1.06. ({This range slightly differs from the previously reported
\citep{PNAS13} range 0.95-1.08 since in the present work the
numerical accuracy of the calculated $\kappa_1$ values for $W
> 100$ has been improved.}) This is understood in the context that
these values correspond to similar critical processes, thus
exhibiting the same dependence of $\beta_W$ on $W$. During the
whole period studied, however, beyond the above mentioned  $\beta$ minima
before all the  ${\rm M}_{JMA} \geq 7.8$ EQs, more minima exist.
Among these minima we choose those which are equally deep or
deeper than the shallowest one of the $\beta_{200,min}$ values
previously identified (e.g., 0.294 in Table \ref{tab2}) and in
addition they have $\beta_{300,min} / \beta_{200,min}$ values
lying in the narrow range determined above. Thus, we now find a
list of ``additional'' $\beta$ minima (see Table \ref{tab3}) that
must be checked whether they are non-precursory or may be followed
by EQs of smaller magnitude.

We now repeat the whole procedure -as described above- for the
determination of $\beta$ minima in the smaller area. Thus, we
obtain a new set of $\beta$ minima (with shallowest
$\beta_{200,min}$=0.293 and $\beta_{300,min} / \beta_{200,min}$
range 0.97-1.09, see Table \ref{tab4}) that appear a few months before all the  ${\rm
M}_{JMA} \geq 7.8$ EQs (cf. there exist 4 such EQs in the smaller
area, see Fig.\ref{fig1}) as well as a new list of ``additional''
$\beta$ minima (see Table \ref{tab5}) to be checked whether they
are non-precursory or may be followed by EQs of smaller magnitude.
Comparing these new $\beta$ minima with the previous ones, we
investigate whether they:

(1) appear practically on the same date (differing by no more than
10 days or so) in both areas, and

(2) exhibit the three main features (i.e., the sequence
{(A)} anti-correlated behavior /
{(B)} correlated /{(C)} almost
random behavior) emerged from the  results of DFA of EQ
magnitude time series discussed in the main text. In Tables
\ref{tab3} and \ref{tab5} corresponding to the ``additional''
minima, we mark in bold the values which do not satisfy at least
one of the three inequalities given below which quantify these
three main features
\[
\alpha_{min,bef} \leq 0.47, \alpha_{corr} > 0.50, \alpha_{min,aft}
\leq 0.50
\]
(Note that considerable errors are introduced in the estimation of the
$\alpha$ exponent when using a relatively small number of points
as the one used here, i.e., 300. This is why we adopt $\alpha =
0.47$ as the maximum value in order to assure anti-correlated
behavior in the period before the appearance of $\beta$ minimum.)

A summary of the main results obtained after carrying out this
investigation is as follows:

First, the $\beta$ minima identified a few months before all the 
${\rm M}_{JMA} \geq 7.8$ EQs with epicenters inside of both areas
obey the aforementioned requirements (1) and (2), see Tables
\ref{tab2} and \ref{tab4}. Remarkably, in the remaining case,
i.e., the East-Off Hokkaido M8.2 EQ labelled EQ2 in Table \ref{tab1},
with an epicenter outside of the smaller area but inside the
larger, we find $\beta$ minimum not only in the study of the
larger area (on 30 June 1994, see Table \ref{tab2}), but also in
the relevant study of the smaller area (see the third $\beta$
minimum in Table \ref{tab5} observed on 5 July 1994). Second, all
the other ``additional'' $\beta$ minima resulting from the studies
in both areas (see Tables \ref{tab3} and \ref{tab5}) violate at
least one of the requirements (1) and (2). In other words, only
the $\beta$ minima appearing a few months before the five  ${\rm
M}_{JMA} \geq 7.8$ EQs exhibit all the aforementioned properties.
Remarkably, this procedure could have been applied before the
occurrence of the M9 Tohoku EQ, after the identification of the
deepest $\beta$ minimum on January 2011 having $\beta_{300,min} /
\beta_{200,min}$ almost unity, thus lying inside the narrow ranges
identified from previous ${\rm M}_{JMA} \geq 7.8$ EQs (see Tables
\ref{tab2} and \ref{tab4}), leading to the conclusion, as mentioned in the main text, that a 
${\rm M}_{JMA} \geq 7.8$ EQ was going to occur in a few months.

We clarify that the above procedure does not preclude of course
that one of the ``additional'' minima may be of truly precursory
nature, but corresponding to an EQ of magnitude smaller than the
threshold adopted. As a first example we mention the case marked
FA7 in Table \ref{tab3} referring to a $\beta_{200,min}$ observed
on 12 April 2000. It preceded the M6.5 EQ that occurred on 1 July
2000 of the volcanic-seismic swarm activity in the Izu Island
region in 2000, see the $\beta_{200}$ curve (red) in Fig.3{C}.
A second example is the case marked EQc in Table \ref{tab3}, which
refers to a $\beta_{200,min}$ on 15 October 1994 that
preceded the M7.6 Far-Off Sanriku EQ
 on 28 December 1994 \citep{PNAS13}.



%
%
%
%
%
%
%

\begin{acknowledgments}
We would like to express our sincere thanks to Professor H. Eugene Stanley, 
Professor Shlomo Havlin and Dr. Joel Tenenbaum for providing us 
the necessary data in order to insert in Fig. \ref{fig1} 
the nodes and the associated links of their network. 
\end{acknowledgments}

\begin{figure}
\noindent\includegraphics[scale=0.85]{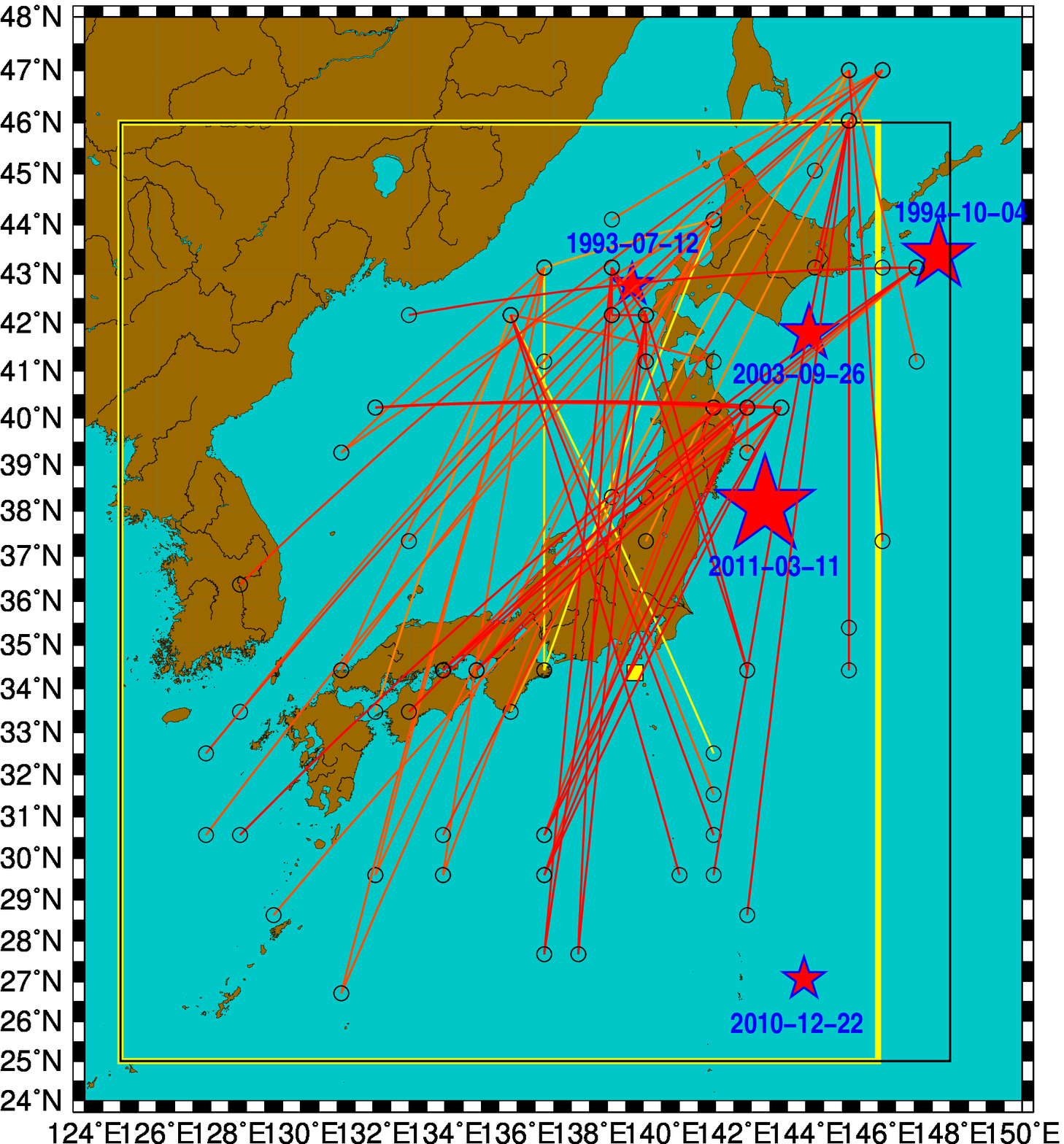}
\caption{(color) The epicenters (stars) of all  EQs with 
magnitude 7.8 or larger within the area ${\rm N}_{25}^{46}$ ${\rm E}_{125}^{148}$ 
(black rectangle) since 1 January 1984 until the M9 Tohoku EQ on 11 March 2011 (Table \ref{tab1}).
The smaller area ${\rm N}_{25}^{46}$ ${\rm
E}_{125}^{146}$ studied by \citet{TECTO12} is also shown with yellow rectangle. 
The small yellow square indicates the location of the Niijima Island 
where the precursory SES activity of the volcanic-seismic 
swarm activity in 2000 in the Izu Island region has been recorded \citep{UYE02,UYE09}.
Furthermore, the network links as reported by 
\citet{TEN12} (see their Fig.6(a)) are also shown.}\label{fig1}
\end{figure}

\begin{figure}
\noindent\includegraphics[scale=0.85]{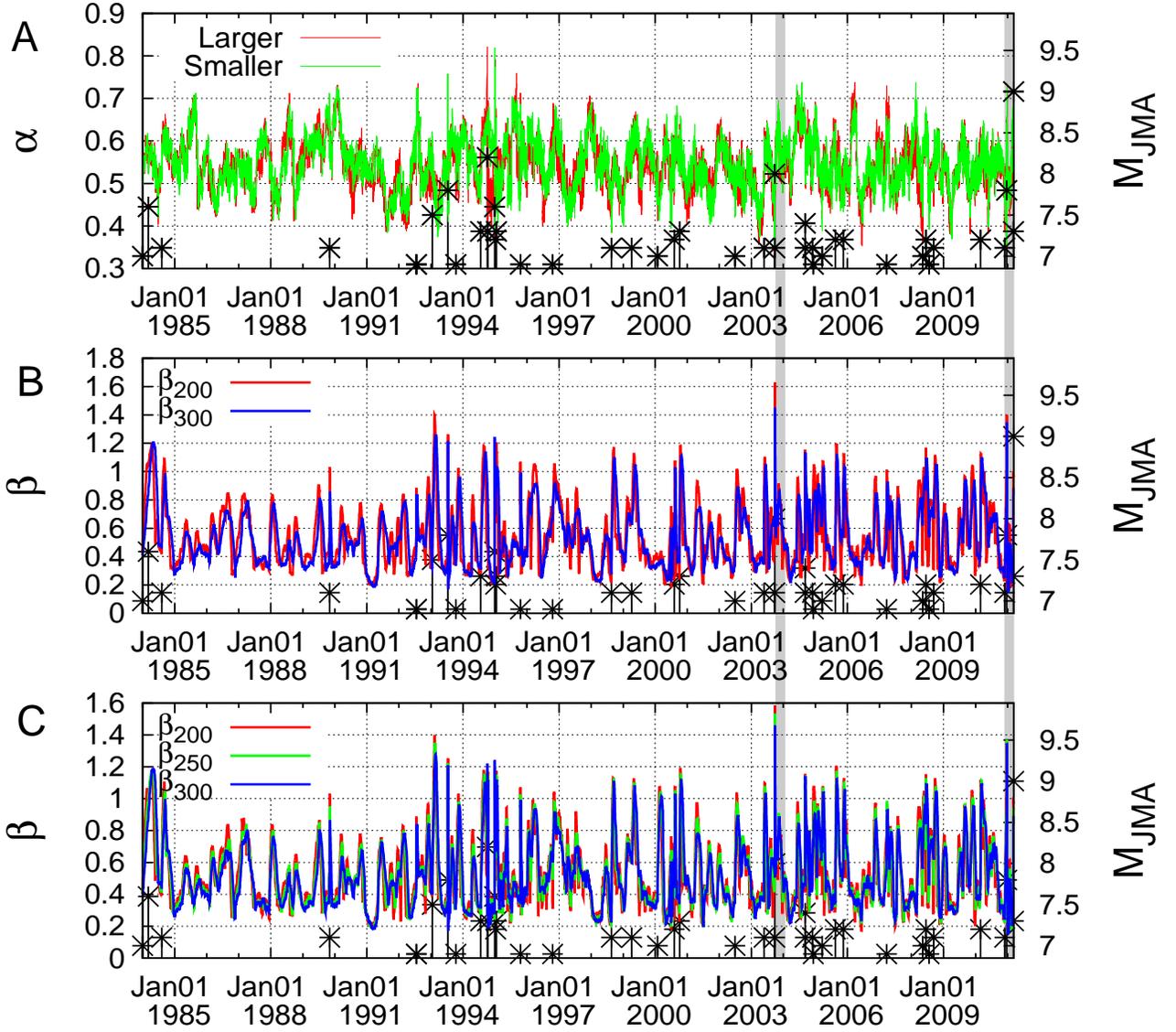}
 \caption{ (color) The DFA exponent $\alpha$ and the variability
$\beta$ of $\kappa_1$ (left scale) of seismicity since 1 January
1984 until just before the M9 Tohoku EQ: ({A}) the exponent
$\alpha$ for $W$=300 in the larger area ${\rm N}_{25}^{46}$ ${\rm
E}_{125}^{148}$ (red) and in the smaller area ${\rm N}_{25}^{46}$
${\rm E}_{125}^{146}$ (green) of Fig.\ref{fig1}. ({B})
The variability $\beta_{200}$ (red) and $\beta_{300}$ (blue) in
the smaller area. ({C}) The variability $\beta_{200}$ (red),
$\beta_{250}$ (green) and $\beta_{300}$ (blue) in the larger area.
In addition, all ${\rm M}_{JMA} \geq 7.0$ EQs (in black,
${\rm M}_{JMA}$ in the right scale) are plotted.}
 \label{fig2}
\end{figure}

\begin{figure}
\noindent\includegraphics[scale=0.85]{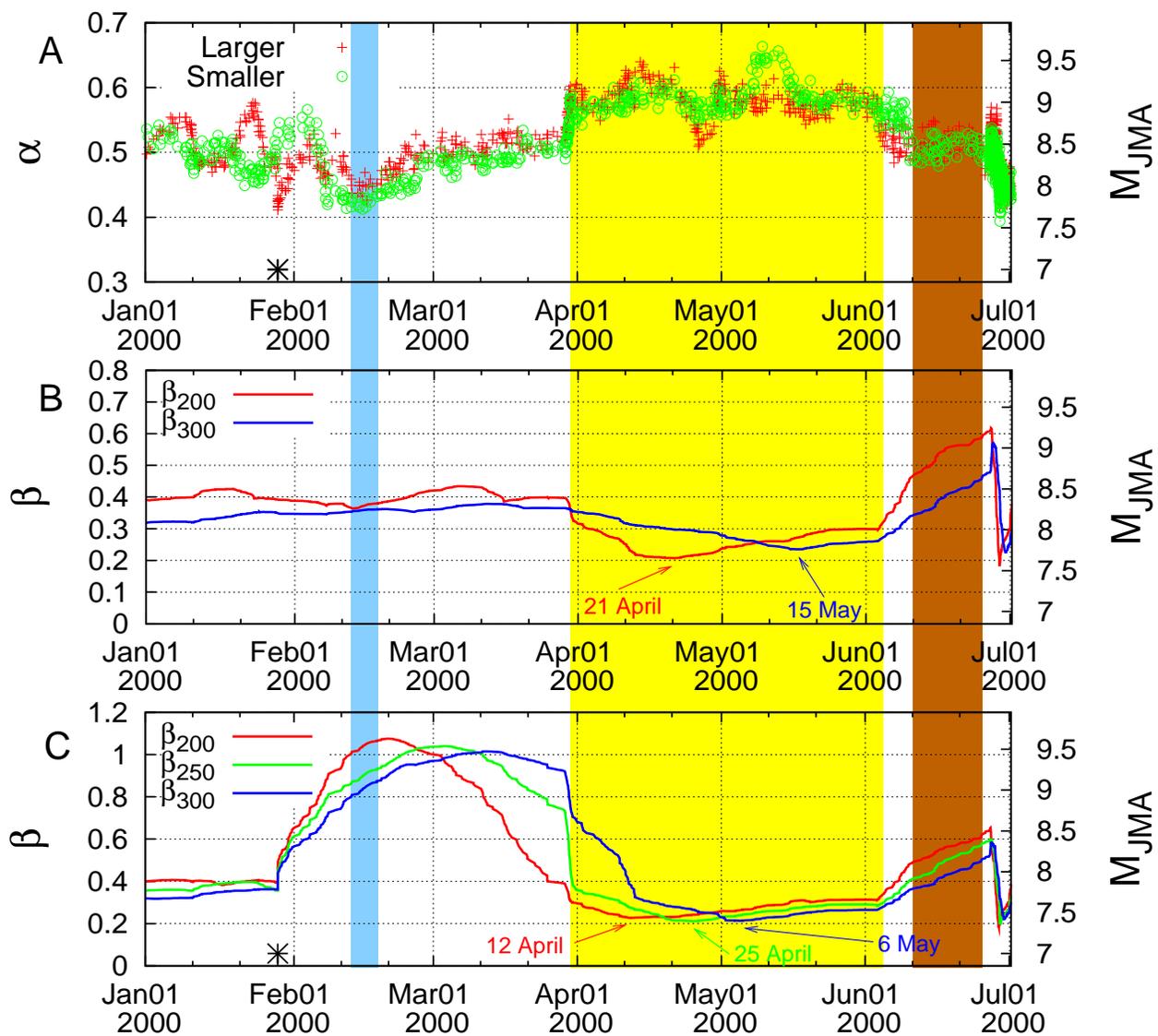}
 \caption{(color) (A)-(C): Excerpt of Fig. \ref{fig2} plotted in expanded 
 time scale from 1 January 2000 until 1 July 2000, which is the date of 
 occurrence of the M6.5 EQ 
 close to Niijima Island during the 
 volcanic-seismic swarm activity in the Izu Island region in 2000.}
 \label{fig3}
\end{figure}

\begin{figure}
\noindent\includegraphics[scale=0.85]{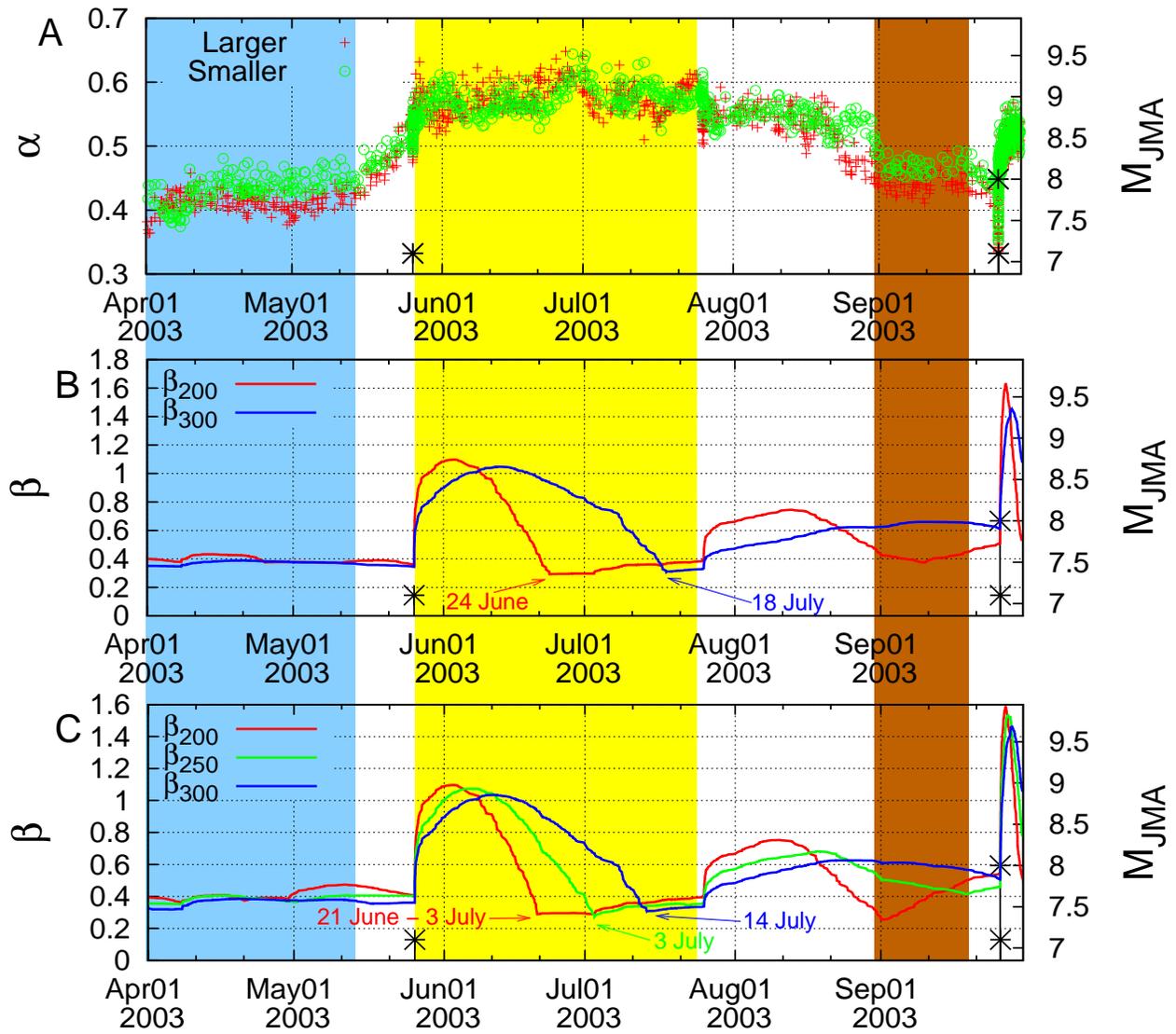}
 \caption{(color) Excerpt of Fig. \ref{fig2}
 plotted in expanded time scale before 
 the M8 Off Tokachi  EQ on 26 September 2003}
 \label{fig4}
\end{figure}

\begin{figure}
\noindent\includegraphics[scale=0.85]{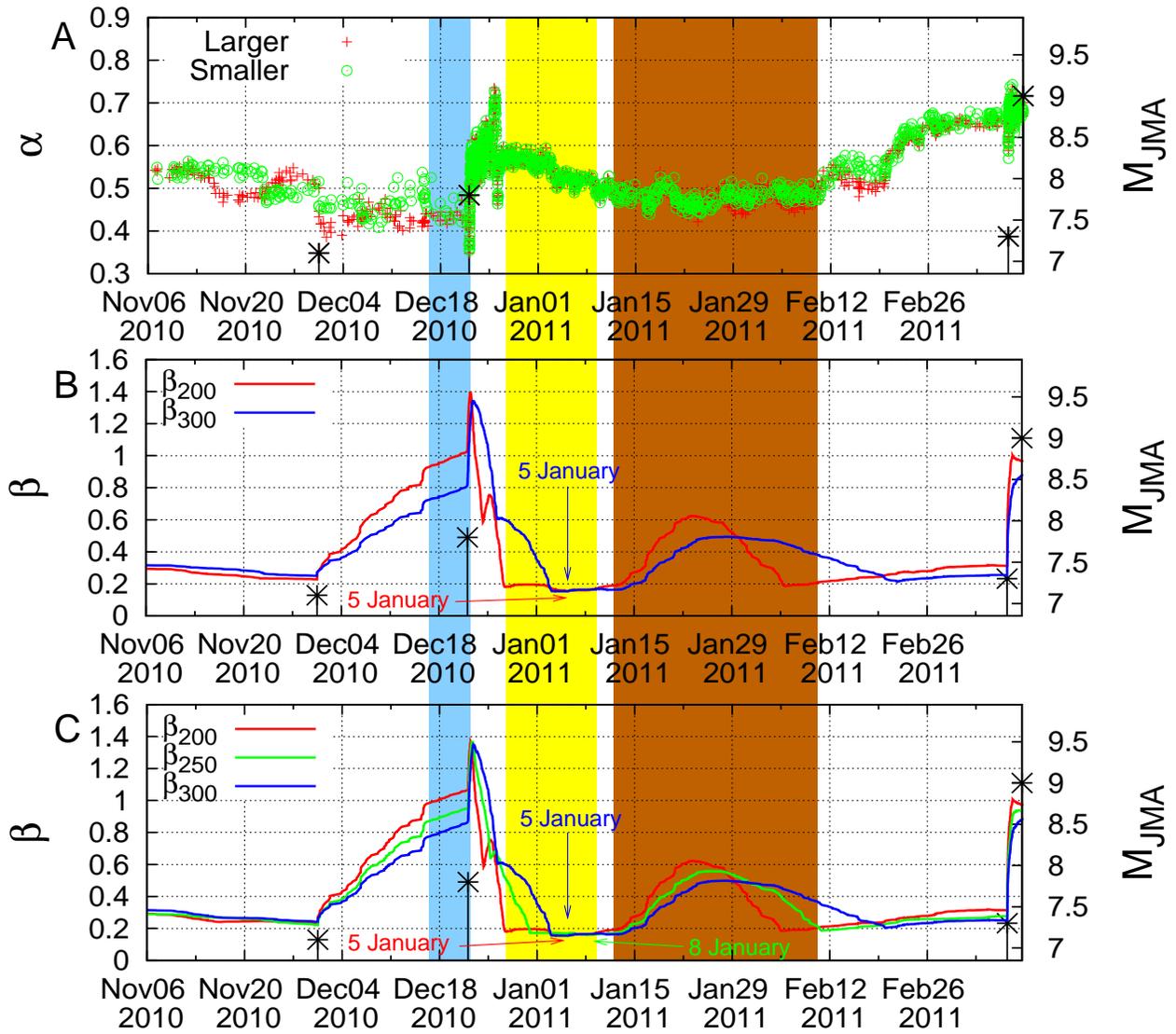}
 \caption{{(color)} Excerpt of Fig. \ref{fig2} plotted 
 in expanded time scale before the M9 Tohoku EQ on 11 March 2011.}
 \label{fig5}
\end{figure}

\begin{figure}
{\includegraphics[scale=0.85]{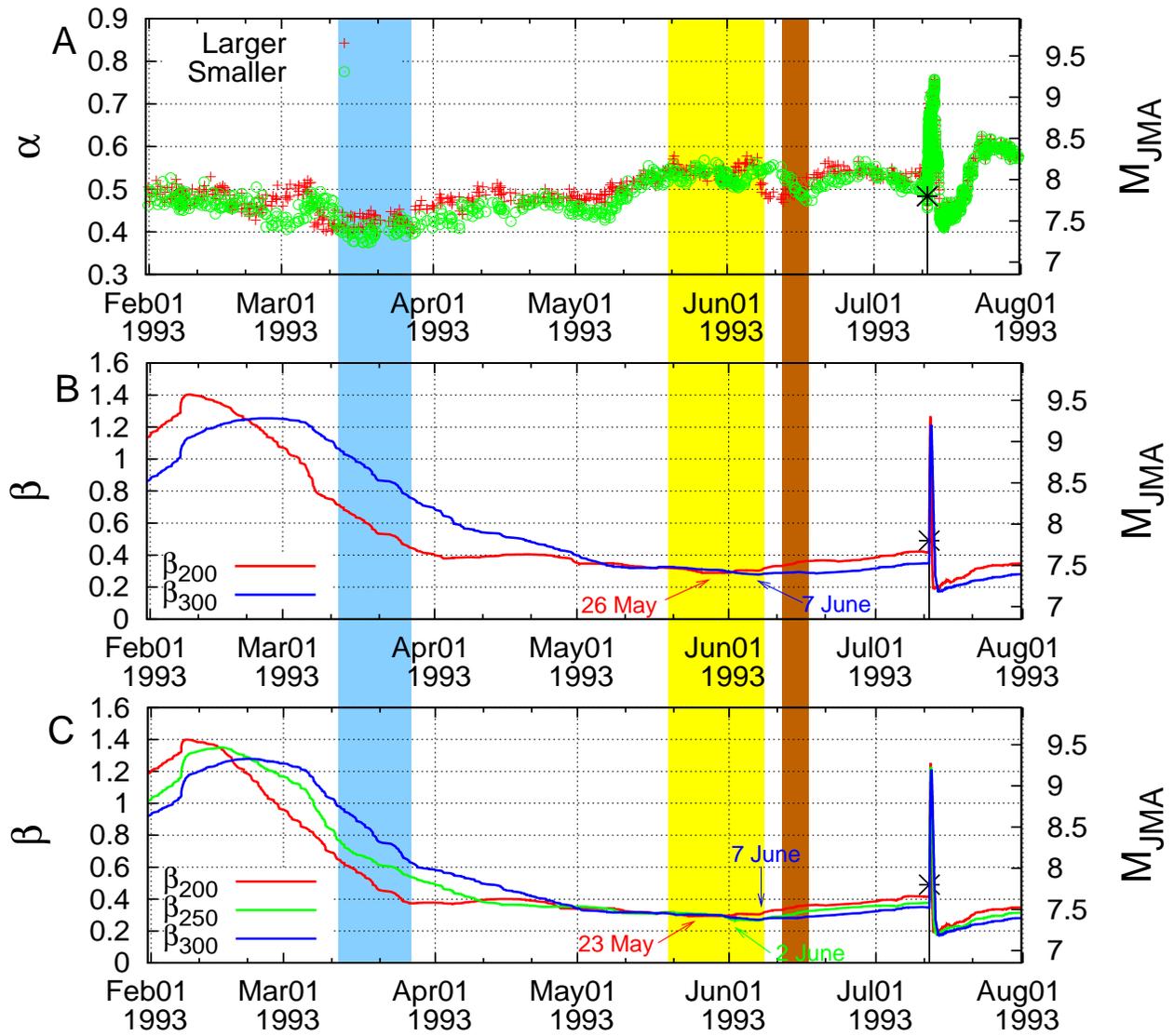}}
\caption{(color) Excerpt of Fig. \ref{fig2}  plotted
in expanded time scale before the M7.8 Southwest-Off Hokkaido EQ
on 12 July 1993. }
\label{figA1}
\end{figure}

\begin{figure}
\noindent\includegraphics[scale=0.85]{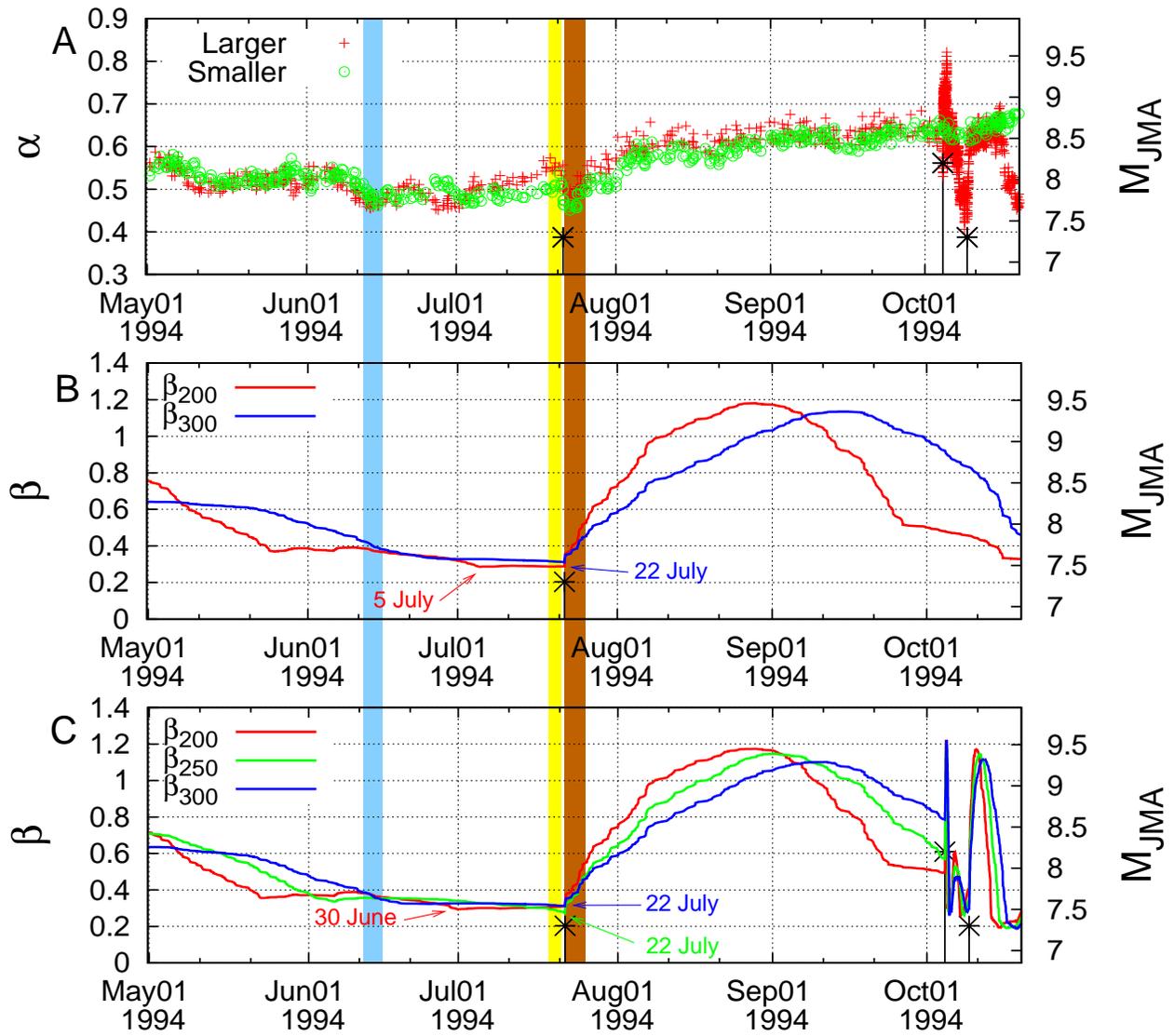}
\caption{(color) Excerpt of Fig. \ref{fig2} plotted
in expanded time scale before the M8.2 East-Off Hokkaido EQ on 4
October 1994. }
\label{figA2}
\end{figure}

\begin{figure}
\noindent\includegraphics[scale=0.85]{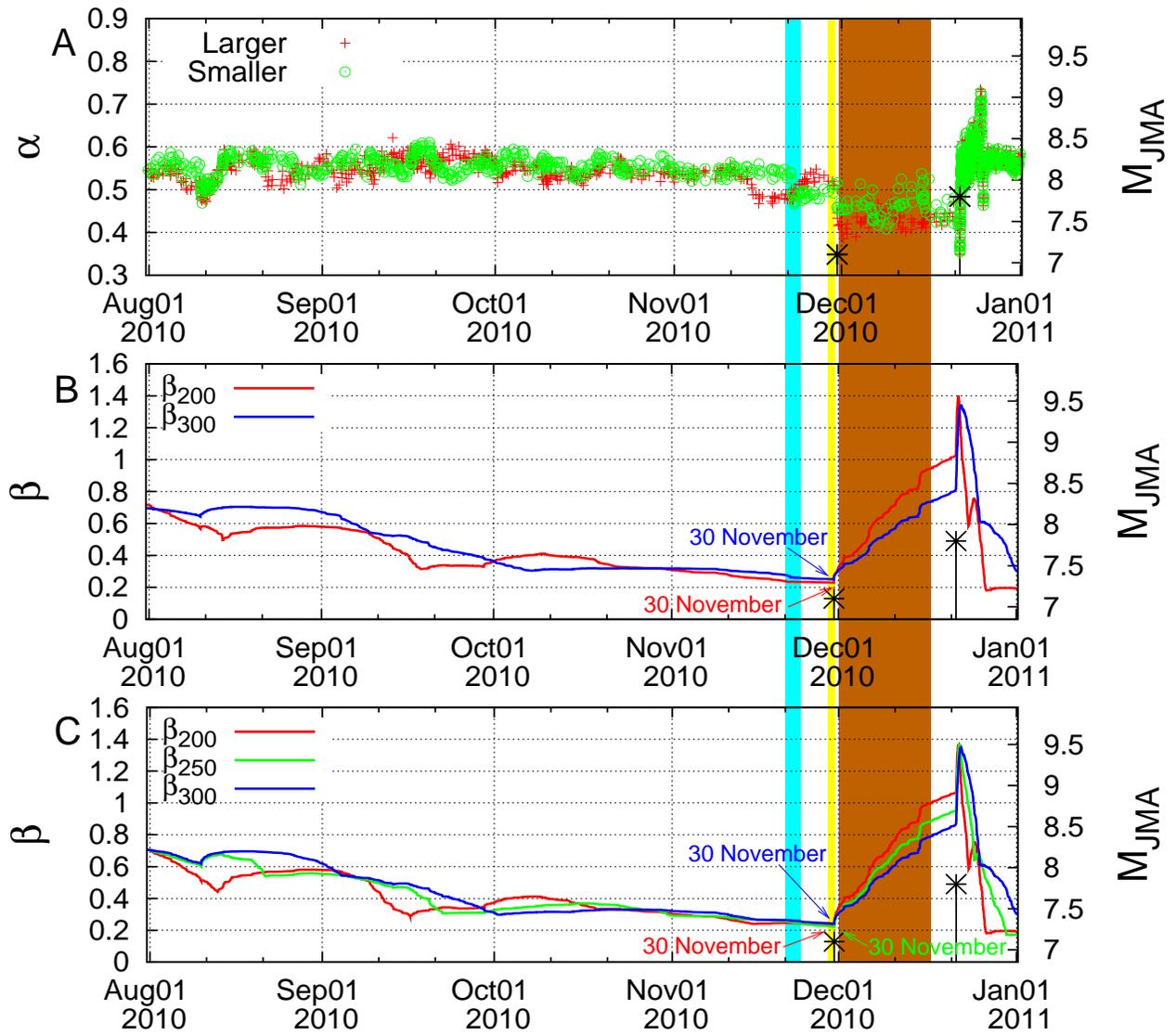}
\caption{{(color)} Excerpt of Fig. \ref{fig2}  plotted
in expanded time scale before the M7.8 Near Chichi-jima EQ on 22
December 2010. }
\label{figA3}
\end{figure}

\clearpage 

\begin{table}
\caption{All  ${\rm M}_{JMA} \geq 7.8$ EQs since 1 January 1984
until the M9 Tohoku EQ on 11 March 2011 in the larger area ${\rm
N}_{25}^{46}  {\rm E}_{125}^{148}$}\label{tab1}
\begin{center}
\begin{tabular}{cccccc}
\hline Label & EQ Date & EQ name & Lat. ($^o$N) & Lon. ($^o$E) & M \\
\hline
EQ1 &        1993-07-12 & Southwest-Off Hokkaido EQ &    42.78 &   139.18 &   7.8 \\
EQ2\footnotemark[1] &        1994-10-04 &
East-Off Hokkaido EQ &    43.38 & 147.67
&   8.2 \\
EQ3 &        2003-09-26 & Off Tokachi EQ &    41.78 &   144.08 &   8.0 \\
EQ4 &        2010-12-22 & Near Chichi-jima EQ &    27.05 &   143.93 &   7.8 \\
EQ5 &        2011-03-11 & Tohoku EQ &    38.10 &   142.86 &   9.0 \\
\hline
\end{tabular}
\end{center}
\footnotemark[1]{This EQ lies outside of the smaller area ${\rm
N}_{25}^{46}  {\rm E}_{125}^{146}$.}
\end{table}

\clearpage
\begin{table}
\caption{The values of $\beta_{W,min}$, $\alpha_{min,bef}$,
$\alpha_{corr}$ and $\alpha_{min,aft}$ in the larger area
investigated with sliding natural time windows of length 6 to $W$
that preceded the ${\rm M}_{JMA} \geq 7.8$ EQs listed in Table
\ref{tab1}. Hereafter, the value of $\alpha_{corr}$ is given when
$\beta_{250,min}$ appears and $\alpha_{min,bef}$ is the minimum of
the DFA exponent up to three and a half months (105 days) before
$\beta_{250,min}$. }\label{tab2}
\begin{center}
\begin{tabular}{cccccccc}
\hline
Label & $\beta_{200,min}$ & $\beta_{250,min}$ & $\beta_{300,min}$ & $\frac{\beta_{300,min}}{\beta_{200,min}}$ & $\alpha_{min,bef}$ & $\alpha_{corr}$ & $\alpha_{min,aft}$ \\
\hline
EQ1 &    0.292(19930523)\footnotemark[1] &    0.266(19930602) &    0.270(19930607) &  0.92 &    0.397(19930327) &    0.515(19930602) &    0.469(19930613) \\
EQ2 &    0.294(19940630) &    0.274(19940722) &    0.312(19940722) &  1.06 &    0.451(19940627) &    0.557(19940722) &    0.476(19940724) \\
EQ3 &    0.289(20030703)\footnotemark[2] &    0.273(20030703) &    0.307(20030714) &  1.06 &    0.356(20030325) &    0.553(20030703) &    0.384(20030925\footnotemark[3]) \\
EQ4 &    0.233(20101130) &    0.221(20101130) &    0.242(20101130) &  1.04 &    0.467(20101116) &    0.501(20101130) &    0.385(20101201) \\
EQ5 &    0.156(20110105) &    0.160(20110108) &    0.154(20110105) &  0.99 &    0.348(20101222) &    0.524(20110108) &    0.422(20110123) \\
\hline
\end{tabular}
\end{center}
\footnotemark[1]{A comparable (within $\pm 0.0005$) value $\beta_{200,min}=$0.293 is also found on 24 May 1993.} \\
\footnotemark[2]{The same (within $\pm 0.0005$) value $\beta_{200,min}=$0.289 is also found on 21 June 2003.}\\
\footnotemark[3]{The date of the M8 EQ is either 25 or 26 September 2003 depending on the use of LT or UT, respectively. Here, we use LT. This value of $\alpha$ is observed 15.5 hours before the mainshock.}
\end{table}
\clearpage

\begin{table}
\caption{The values of $\beta_{W,min}$, $\alpha_{min,bef}$,
$\alpha_{corr}$ and $\alpha_{min,aft}$ for the ``additional''
minima identified in the larger area investigated with sliding
natural time windows of length 6 to $W$ }\label{tab3}
\begin{center}
\begin{tabular}{cccccccc}
\hline
Label & $\beta_{200,min}$ & $\beta_{250,min}$ & $\beta_{300,min}$ & $\frac{\beta_{300,min}}{\beta_{200,min}}$ & $\alpha_{min,bef}$ & $\alpha_{corr}$ & $\alpha_{min,aft}$ \\
\hline
FA1 &    0.253(19861013) &    0.256(19861028) &
0.246(19861115) &  0.97 &    {\bf 0.484(19860905)}\footnotemark[1]                                    &    0.557(19861028) &    0.446(19861129) \\
FA2 &    0.279(19890808) &    0.282(19890826) &    0.285(19890915) &  1.02 &    {\bf 0.524(19890730)} &    0.566(19890826) &    {\bf 0.508(19891128)} \\
FA3 &    0.251(19920405) &    0.242(19920423) &    0.247(19920510) &  0.98 &    0.389(19920325)       &    {\bf 0.411(19920423)} &    0.411(19920423) \\
FA4 &    0.187(19930713) &    0.176(19930714) &    0.174(19930715) &  0.93 &    0.441(19930407)       &    0.584(19930714) &    0.410(19930716) \\
EQc\footnotemark[2]
    &    0.194(19941015) &    0.191(19941017) &    0.188(19941019) &  0.97 &    0.406(19941009)       &    0.584(19941017) &    0.399(19941125) \\
FA5 &    0.236(19980217) &    0.214(19980228) &    0.223(19980312) &  0.95 &    {\bf 0.524(19980211)} &    0.540(19980228) &    0.470(19980509) \\
FA6 &    0.281(19981218) &    0.299(19981230) &    0.294(19990110) &  1.05 &    0.423(19981223)       &    {\bf 0.474(19981230)} &    0.457(19990108) \\
FA7 &    0.227(20000412)\footnotemark[3] &    0.211(20000425) &    0.213(20000506) &  0.94 &    0.411(20000129)\footnotemark[4]       &    0.571(20000425) &    0.411(20000702) \\
FA8 &    0.243(20020512) &    0.231(20020523) &    0.253(20020603) &  1.04 &    {\bf 0.495(20020515)} &    0.527(20020523) &    0.424(20020630) \\
FA9 &    0.233(20040221) &    0.224(20040302) &    0.219(20040315) &  0.94 &    {\bf 0.478(20031224)} &    0.512(20040302) &    0.447(20040319) \\
FA10 &   0.283(20050611) &    0.305(20050620) &    0.300(20050701) &  1.06 &    0.413(20050320)       &    {\bf 0.460(20050620)} &    0.430(20050620) \\
FA11 &   0.267(20080227) &    0.287(20080227) &    0.284(20080227) &  1.06 &    {\bf 0.507(20080103)} &    0.668(20080227) &    0.360(20080508) \\
\hline
\end{tabular}
\end{center}
\footnotemark[1]{The values shown in bold violate at least one of the main three features $\alpha_{min,bef} \leq 0.47$, $\alpha_{corr} > 0.50$, $\alpha_{min,aft} \leq 0.50$}              \\
\footnotemark[2]{This minimum was followed by the M7.6 Far-Off Sanriku EQ on 28 December 1994, see \citet{PNAS13}} \\
\footnotemark[3]{The same (within $\pm 0.0005$) value $\beta_{200,min}=$0.227 is also found on 13 April 2000.} \\
\footnotemark[4]{This value and date of $\alpha_{min,bef}$ results of course when {\em solely} 
considering the DFA exponent in the larger area. Upon demanding however a common behavior of the DFA
exponent in {\em both} areas, we find the result described in the text which is marked in Fig.\ref{fig3} in cyan.}
\end{table}

\clearpage

\begin{table}
\caption[]{The values of $\beta_{W,min}$, $\alpha_{min,bef}$,
$\alpha_{corr}$ and $\alpha_{min,aft}$ in the smaller area ${\rm
N}_{25}^{46}  {\rm E}_{125}^{146}$ investigated with sliding
natural time windows of length 6 to $W$ that preceded ${\rm
M}_{JMA} \geq 7.8$ EQs.}\label{tab4}
\begin{center}
\begin{tabular}{cccccccc}
\hline
Label & $\beta_{200,min}$ & $\beta_{250,min}$ & $\beta_{300,min}$ & $\frac{\beta_{300,min}}{\beta_{200,min}}$ & $\alpha_{min,bef}$ & $\alpha_{corr}$ & $\alpha_{min,aft}$ \\
\hline
EQ1 &    0.288(19930526) &    0.264(19930607) &    0.278(19930607) &  0.97 &    0.374(19930319) &    0.557(19930607) &    0.473(19930617) \\
EQ3\footnotemark[1] &    0.293(20030624) &    0.311(20030709) &    0.309(20030718) &  1.06 &    0.374(20030408) &    0.570(20030709) &    0.434(20030925\footnotemark[2]) \\
EQ4 &    0.228(20101130) &    0.233(20101130) &    0.250(20101130) &  1.09 &    0.467(20101122) &    0.516(20101130) &    0.408(20101209) \\
EQ5 &    0.156(20110105) &    0.160(20110108) &    0.154(20110105) &  0.99 &    0.354(20101222) &    0.522(20110108) &    0.439(20110117) \\
\hline
\end{tabular}
\end{center}
\footnotemark[1]{The epicenter of EQ2 -not included here- lies outside of the smaller area.} \\
\footnotemark[2]{The date of the M8 EQ is either 25 or 26 September 2003 depending on the use of LT or UT, respectively. Here, we use LT. This value of $\alpha$ is observed 24.5 hours before the mainshock.}
\end{table}

\clearpage

\begin{table}
\caption{The values of $\beta_{W,min}$, $\alpha_{min,bef}$,
$\alpha_{corr}$ and $\alpha_{min,aft}$ for the ``additional''
minima identified in the smaller area investigated with sliding
natural time windows of length 6 to $W$}\label{tab5}
\begin{center}
\begin{tabular}{cccccccc}
\hline
Label & $\beta_{200,min}$ & $\beta_{250,min}$ & $\beta_{300,min}$ & $\frac{\beta_{300,min}}{\beta_{200,min}}$ & $\alpha_{min,bef}$ & $\alpha_{corr}$ & $\alpha_{min,aft}$ \\
\hline
Fa1 &    0.276(19890810) &    0.278(19890830) &
0.294(19890921) &  1.06 &    {\bf 0.516(19890805)}\footnotemark[1]
                                                                                                &    0.589(19890830) &    {\bf 0.513(19891129)} \\
Fa2 &    0.263(19920412)\footnotemark[2]
                         &    0.261(19920503) &    0.264(19920520) &  1.00 &    0.409(19920330) &    {\bf 0.474(19920503)} &    0.465(19920507) \\
Fa3\footnotemark[3]
    &    0.286(19940705) &    0.297(19940722) &    0.310(19940722) &  1.09 &    0.461(19940615) &    0.501(19940722) &    0.451(19940723) \\
EQc\footnotemark[4]
 &    0.205(19941128)\footnotemark[5]
                         &    0.211(19941210) &    0.212(19941227) &  1.03 &    {\bf 0.539(19941109)} &    0.615(19941210) &    0.478(19941228) \\
Fa4 &    0.252(19950511)\footnotemark[6]
                         &    0.238(19950519) &    0.262(19950528) &  1.04 &    0.441(19950204) &    0.566(19950519) &    0.436(19950702) \\
Fa5 &    0.220(19980227) &    0.230(19980313) &    0.222(19980330) &  1.01 &    {\bf 0.481(19980301)} &    0.536(19980313) &    0.490(19980423) \\
Fa6 &    0.238(20070701) &    0.249(20070701) &    0.240(20070701) &  1.01 &    0.437(20070609) &    0.523(20070701) &    0.454(20070720) \\
\hline
\end{tabular}
\end{center}
\end{table}
\footnotemark[1]{The
values shown in bold violate at least one of the main three
features $\alpha_{min,bef} \leq 0.47$, $\alpha_{corr} > 0.50$,
$\alpha_{min,aft} \leq 0.50$} \\
\footnotemark[2]{The same (within $\pm 0.0005$)
value $\beta_{200,min}=$0.263 is also found on 13 April 1992.} \\
\footnotemark[3]{This minimum was followed by EQ2 which has an
epicenter outside of the smaller area (but inside the larger).} \\
\footnotemark[4]{This minimum was followed by the M7.6 Far-Off
Sanriku EQ on 28 December 1994, see \citet{PNAS13}.} \\
\footnotemark[5]{The same (within $\pm 0.0005$) value $\beta_{200,min}=$0.205 is also found on 30 November 1994.} \\
\footnotemark[6]{The same (within $\pm 0.0005$)
value $\beta_{200,min}=$0.252 is also found on 12 May 1995.}

%
%



\begin{thebibliography}{69}
\expandafter\ifx\csname natexlab\endcsname\relax\def\natexlab#1{#1}\fi
\expandafter\ifx\csname bibnamefont\endcsname\relax
  \def\bibnamefont#1{#1}\fi
\expandafter\ifx\csname bibfnamefont\endcsname\relax
  \def\bibfnamefont#1{#1}\fi
\expandafter\ifx\csname citenamefont\endcsname\relax
  \def\citenamefont#1{#1}\fi
\expandafter\ifx\csname url\endcsname\relax
  \def\url#1{\texttt{#1}}\fi
\expandafter\ifx\csname urlprefix\endcsname\relax\def\urlprefix{URL }\fi
\providecommand{\bibinfo}[2]{#2}
\providecommand{\eprint}[2][]{\url{#2}}

\bibitem[{\citenamefont{Varotsos and Alexopoulos}(1984)}]{VAR84A}
\bibinfo{author}{\bibfnamefont{P.}~\bibnamefont{Varotsos}} \bibnamefont{and}
  \bibinfo{author}{\bibfnamefont{K.}~\bibnamefont{Alexopoulos}},
  \bibinfo{journal}{Tectonophysics} \textbf{\bibinfo{volume}{110}},
  \bibinfo{pages}{73} (\bibinfo{year}{1984}).

\bibitem[{\citenamefont{Varotsos et~al.}(1996)\citenamefont{Varotsos, Eftaxias,
  Lazaridou, Nomicos, Sarlis, Bogris, Makris, Antonopoulos, and
  Kopanas}}]{VAR96B}
\bibinfo{author}{\bibfnamefont{P.}~\bibnamefont{Varotsos}},
  \bibinfo{author}{\bibfnamefont{K.}~\bibnamefont{Eftaxias}},
  \bibinfo{author}{\bibfnamefont{M.}~\bibnamefont{Lazaridou}},
  \bibinfo{author}{\bibfnamefont{K.}~\bibnamefont{Nomicos}},
  \bibinfo{author}{\bibfnamefont{N.}~\bibnamefont{Sarlis}},
  \bibinfo{author}{\bibfnamefont{N.}~\bibnamefont{Bogris}},
  \bibinfo{author}{\bibfnamefont{J.}~\bibnamefont{Makris}},
  \bibinfo{author}{\bibfnamefont{G.}~\bibnamefont{Antonopoulos}},
  \bibnamefont{and} \bibinfo{author}{\bibfnamefont{J.}~\bibnamefont{Kopanas}},
  \bibinfo{journal}{Acta Geophysica Polonica} \textbf{\bibinfo{volume}{44}},
  \bibinfo{pages}{301} (\bibinfo{year}{1996}).

\bibitem[{\citenamefont{Varotsos and Alexopoulos}(1986)}]{VARBOOK}
\bibinfo{author}{\bibfnamefont{P.}~\bibnamefont{Varotsos}} \bibnamefont{and}
  \bibinfo{author}{\bibfnamefont{K.}~\bibnamefont{Alexopoulos}},
  \emph{\bibinfo{title}{Thermodynamics of Point Defects and their Relation with
  Bulk Properties}} (\bibinfo{publisher}{North Holland},
  \bibinfo{address}{Amsterdam}, \bibinfo{year}{1986}).

\bibitem[{\citenamefont{Varotsos et~al.}(1993)\citenamefont{Varotsos,
  Alexopoulos, and Lazaridou}}]{VAR93}
\bibinfo{author}{\bibfnamefont{P.}~\bibnamefont{Varotsos}},
  \bibinfo{author}{\bibfnamefont{K.}~\bibnamefont{Alexopoulos}},
  \bibnamefont{and}
  \bibinfo{author}{\bibfnamefont{M.}~\bibnamefont{Lazaridou}},
  \bibinfo{journal}{Tectonophysics} \textbf{\bibinfo{volume}{224}},
  \bibinfo{pages}{1} (\bibinfo{year}{1993}).

\bibitem[{\citenamefont{Varotsos and Miliotis}(1974)}]{VAROTSOS1974927}
\bibinfo{author}{\bibfnamefont{P.}~\bibnamefont{Varotsos}} \bibnamefont{and}
  \bibinfo{author}{\bibfnamefont{D.}~\bibnamefont{Miliotis}},
  \bibinfo{journal}{Journal of Physics and Chemistry of Solids}
  \textbf{\bibinfo{volume}{35}}, \bibinfo{pages}{927 } (\bibinfo{year}{1974}).

\bibitem[{\citenamefont{Kostopoulos et~al.}(1975)\citenamefont{Kostopoulos,
  Varotsos, and Mourikis}}]{KOS75}
\bibinfo{author}{\bibfnamefont{D.}~\bibnamefont{Kostopoulos}},
  \bibinfo{author}{\bibfnamefont{P.}~\bibnamefont{Varotsos}}, \bibnamefont{and}
  \bibinfo{author}{\bibfnamefont{S.}~\bibnamefont{Mourikis}},
  \bibinfo{journal}{Can. J. Phys.} \textbf{\bibinfo{volume}{53}},
  \bibinfo{pages}{1318} (\bibinfo{year}{1975}).

\bibitem[{\citenamefont{Varotsos and Alexopoulos}(1977)}]{VAR997}
\bibinfo{author}{\bibfnamefont{P.}~\bibnamefont{Varotsos}} \bibnamefont{and}
  \bibinfo{author}{\bibfnamefont{K.}~\bibnamefont{Alexopoulos}},
  \bibinfo{journal}{J Phys. Chem. Solids} \textbf{\bibinfo{volume}{38}},
  \bibinfo{pages}{997 } (\bibinfo{year}{1977}).

\bibitem[{\citenamefont{Varotsos and Alexopoulos}(1980)}]{VARALEX80p}
\bibinfo{author}{\bibfnamefont{P.}~\bibnamefont{Varotsos}} \bibnamefont{and}
  \bibinfo{author}{\bibfnamefont{K.}~\bibnamefont{Alexopoulos}},
  \bibinfo{journal}{Phys. Rev. B} \textbf{\bibinfo{volume}{21}},
  \bibinfo{pages}{4898} (\bibinfo{year}{1980}).

\bibitem[{\citenamefont{Varotsos and Alexopoulos}(1978)}]{VAR78}
\bibinfo{author}{\bibfnamefont{P.}~\bibnamefont{Varotsos}} \bibnamefont{and}
  \bibinfo{author}{\bibfnamefont{K.}~\bibnamefont{Alexopoulos}},
  \bibinfo{journal}{Phys. Status Solidi A} \textbf{\bibinfo{volume}{47}},
  \bibinfo{pages}{K133} (\bibinfo{year}{1978}).

\bibitem[{\citenamefont{Varotsos}(1980)}]{VAR80K133}
\bibinfo{author}{\bibfnamefont{P.}~\bibnamefont{Varotsos}},
  \bibinfo{journal}{physica status solidi (b)} \textbf{\bibinfo{volume}{100}},
  \bibinfo{pages}{K133} (\bibinfo{year}{1980}).

\bibitem[{\citenamefont{Varotsos and Alexopoulos}(1982)}]{VARALEX82B}
\bibinfo{author}{\bibfnamefont{P.}~\bibnamefont{Varotsos}} \bibnamefont{and}
  \bibinfo{author}{\bibfnamefont{K.}~\bibnamefont{Alexopoulos}},
  \bibinfo{journal}{physica status solidi (b)} \textbf{\bibinfo{volume}{110}},
  \bibinfo{pages}{9} (\bibinfo{year}{1982}).

\bibitem[{\citenamefont{Varotsos et~al.}(1982)\citenamefont{Varotsos,
  Alexopoulos, and Nomicos}}]{VARALEX82}
\bibinfo{author}{\bibfnamefont{P.}~\bibnamefont{Varotsos}},
  \bibinfo{author}{\bibfnamefont{K.}~\bibnamefont{Alexopoulos}},
  \bibnamefont{and} \bibinfo{author}{\bibfnamefont{K.}~\bibnamefont{Nomicos}},
  \bibinfo{journal}{Phys. Status Solidi B} \textbf{\bibinfo{volume}{111}},
  \bibinfo{pages}{581} (\bibinfo{year}{1982}).

\bibitem[{\citenamefont{Varotsos}(2008)}]{VAR08438}
\bibinfo{author}{\bibfnamefont{P.}~\bibnamefont{Varotsos}},
  \bibinfo{journal}{Solid State Ionics} \textbf{\bibinfo{volume}{179}},
  \bibinfo{pages}{438 } (\bibinfo{year}{2008}).

\bibitem[{\citenamefont{Varotsos
  et~al.}(2003{\natexlab{a}})\citenamefont{Varotsos, Sarlis, and
  Skordas}}]{NAT03A}
\bibinfo{author}{\bibfnamefont{P.~A.} \bibnamefont{Varotsos}},
  \bibinfo{author}{\bibfnamefont{N.~V.} \bibnamefont{Sarlis}},
  \bibnamefont{and} \bibinfo{author}{\bibfnamefont{E.~S.}
  \bibnamefont{Skordas}}, \bibinfo{journal}{Phys. Rev. E}
  \textbf{\bibinfo{volume}{67}}, \bibinfo{pages}{021109}
  (\bibinfo{year}{2003}{\natexlab{a}}).

\bibitem[{\citenamefont{Varotsos
  et~al.}(2003{\natexlab{b}})\citenamefont{Varotsos, Sarlis, and
  Skordas}}]{NAT03B}
\bibinfo{author}{\bibfnamefont{P.~A.} \bibnamefont{Varotsos}},
  \bibinfo{author}{\bibfnamefont{N.~V.} \bibnamefont{Sarlis}},
  \bibnamefont{and} \bibinfo{author}{\bibfnamefont{E.~S.}
  \bibnamefont{Skordas}}, \bibinfo{journal}{Phys. Rev. E}
  \textbf{\bibinfo{volume}{68}}, \bibinfo{pages}{031106}
  (\bibinfo{year}{2003}{\natexlab{b}}).

\bibitem[{\citenamefont{Varotsos et~al.}(2008)\citenamefont{Varotsos, Sarlis,
  Skordas, and Lazaridou}}]{NAT08}
\bibinfo{author}{\bibfnamefont{P.~A.} \bibnamefont{Varotsos}},
  \bibinfo{author}{\bibfnamefont{N.~V.} \bibnamefont{Sarlis}},
  \bibinfo{author}{\bibfnamefont{E.~S.} \bibnamefont{Skordas}},
  \bibnamefont{and} \bibinfo{author}{\bibfnamefont{M.~S.}
  \bibnamefont{Lazaridou}}, \bibinfo{journal}{J. Appl. Phys.}
  \textbf{\bibinfo{volume}{103}}, \bibinfo{eid}{014906} (\bibinfo{year}{2008}).

\bibitem[{\citenamefont{Varotsos et~al.}(2009)\citenamefont{Varotsos, Sarlis,
  and Skordas}}]{NAT09V}
\bibinfo{author}{\bibfnamefont{P.~A.} \bibnamefont{Varotsos}},
  \bibinfo{author}{\bibfnamefont{N.~V.} \bibnamefont{Sarlis}},
  \bibnamefont{and} \bibinfo{author}{\bibfnamefont{E.~S.}
  \bibnamefont{Skordas}}, \bibinfo{journal}{CHAOS}
  \textbf{\bibinfo{volume}{19}}, \bibinfo{pages}{023114}
  (\bibinfo{year}{2009}).

\bibitem[{\citenamefont{Ren et~al.}(2012)\citenamefont{Ren, Chen, and
  Huang}}]{REN12}
\bibinfo{author}{\bibfnamefont{H.}~\bibnamefont{Ren}},
  \bibinfo{author}{\bibfnamefont{X.}~\bibnamefont{Chen}}, \bibnamefont{and}
  \bibinfo{author}{\bibfnamefont{Q.}~\bibnamefont{Huang}},
  \bibinfo{journal}{Geophys. J. Int.} \textbf{\bibinfo{volume}{188}},
  \bibinfo{pages}{925} (\bibinfo{year}{2012}),
  \eprint{http://gji.oxfordjournals.org/content/188/3/925.full.pdf+html}.

\bibitem[{\citenamefont{Varotsos}(1978)}]{VARO78}
\bibinfo{author}{\bibfnamefont{P.}~\bibnamefont{Varotsos}},
  \bibinfo{journal}{Phys. Stat. Sol. (b)} \textbf{\bibinfo{volume}{90}},
  \bibinfo{pages}{339} (\bibinfo{year}{1978}).

\bibitem[{\citenamefont{Huang}(2011)}]{HUA11A}
\bibinfo{author}{\bibfnamefont{Q.}~\bibnamefont{Huang}}, \bibinfo{journal}{Nat.
  Hazards Earth Syst. Sci.} \textbf{\bibinfo{volume}{11}},
  \bibinfo{pages}{2941} (\bibinfo{year}{2011}).

\bibitem[{\citenamefont{Varotsos
  et~al.}(2011{\natexlab{a}})\citenamefont{Varotsos, Sarlis, and
  Skordas}}]{SPRINGER}
\bibinfo{author}{\bibfnamefont{P.~A.} \bibnamefont{Varotsos}},
  \bibinfo{author}{\bibfnamefont{N.~V.} \bibnamefont{Sarlis}},
  \bibnamefont{and} \bibinfo{author}{\bibfnamefont{E.~S.}
  \bibnamefont{Skordas}}, \emph{\bibinfo{title}{Natural Time Analysis: The new
  view of time. Precursory Seismic Electric Signals, Earthquakes and other
  Complex Time-Series}} (\bibinfo{publisher}{Springer-Verlag},
  \bibinfo{address}{Berlin Heidelberg}, \bibinfo{year}{2011}{\natexlab{a}}).

\bibitem[{\citenamefont{Varotsos and Lazaridou}(1991)}]{VAR91}
\bibinfo{author}{\bibfnamefont{P.}~\bibnamefont{Varotsos}} \bibnamefont{and}
  \bibinfo{author}{\bibfnamefont{M.}~\bibnamefont{Lazaridou}},
  \bibinfo{journal}{Tectonophysics} \textbf{\bibinfo{volume}{188}},
  \bibinfo{pages}{321} (\bibinfo{year}{1991}).

\bibitem[{\citenamefont{Uyeda et~al.}(2000)\citenamefont{Uyeda, Nagao, Orihara,
  Yamaguchi, and Takahashi}}]{UYE00}
\bibinfo{author}{\bibfnamefont{S.}~\bibnamefont{Uyeda}},
  \bibinfo{author}{\bibfnamefont{T.}~\bibnamefont{Nagao}},
  \bibinfo{author}{\bibfnamefont{Y.}~\bibnamefont{Orihara}},
  \bibinfo{author}{\bibfnamefont{T.}~\bibnamefont{Yamaguchi}},
  \bibnamefont{and}
  \bibinfo{author}{\bibfnamefont{I.}~\bibnamefont{Takahashi}},
  \bibinfo{journal}{Proc. Natl. Acad. Sci. USA} \textbf{\bibinfo{volume}{97}},
  \bibinfo{pages}{4561} (\bibinfo{year}{2000}).

\bibitem[{\citenamefont{Uyeda et~al.}(2002)\citenamefont{Uyeda, Hayakawa,
  Nagao, Molchanov, Hattori, Orihara, Gotoh, Akinaga, and Tanaka}}]{UYE02}
\bibinfo{author}{\bibfnamefont{S.}~\bibnamefont{Uyeda}},
  \bibinfo{author}{\bibfnamefont{M.}~\bibnamefont{Hayakawa}},
  \bibinfo{author}{\bibfnamefont{T.}~\bibnamefont{Nagao}},
  \bibinfo{author}{\bibfnamefont{O.}~\bibnamefont{Molchanov}},
  \bibinfo{author}{\bibfnamefont{K.}~\bibnamefont{Hattori}},
  \bibinfo{author}{\bibfnamefont{Y.}~\bibnamefont{Orihara}},
  \bibinfo{author}{\bibfnamefont{K.}~\bibnamefont{Gotoh}},
  \bibinfo{author}{\bibfnamefont{Y.}~\bibnamefont{Akinaga}}, \bibnamefont{and}
  \bibinfo{author}{\bibfnamefont{H.}~\bibnamefont{Tanaka}},
  \bibinfo{journal}{Proc. Natl. Acad. Sci. USA} \textbf{\bibinfo{volume}{99}},
  \bibinfo{pages}{7352} (\bibinfo{year}{2002}).

\bibitem[{\citenamefont{Uyeda et~al.}(2009)\citenamefont{Uyeda, Kamogawa, and
  Tanaka}}]{UYE09}
\bibinfo{author}{\bibfnamefont{S.}~\bibnamefont{Uyeda}},
  \bibinfo{author}{\bibfnamefont{M.}~\bibnamefont{Kamogawa}}, \bibnamefont{and}
  \bibinfo{author}{\bibfnamefont{H.}~\bibnamefont{Tanaka}},
  \bibinfo{journal}{J. Geophys. Res.} \textbf{\bibinfo{volume}{114}},
  \bibinfo{pages}{B02310} (\bibinfo{year}{2009}).

\bibitem[{\citenamefont{Orihara et~al.}(2012)\citenamefont{Orihara, Kamogawa,
  Nagao, and Uyeda}}]{ORI12}
\bibinfo{author}{\bibfnamefont{Y.}~\bibnamefont{Orihara}},
  \bibinfo{author}{\bibfnamefont{M.}~\bibnamefont{Kamogawa}},
  \bibinfo{author}{\bibfnamefont{T.}~\bibnamefont{Nagao}}, \bibnamefont{and}
  \bibinfo{author}{\bibfnamefont{S.}~\bibnamefont{Uyeda}},
  \bibinfo{journal}{Proc. Natl. Acad. Sci. U.S.A.}
  \textbf{\bibinfo{volume}{109}}, \bibinfo{pages}{19125}
  (\bibinfo{year}{2012}).

\bibitem[{\citenamefont{Turcotte}(1997)}]{TUR97}
\bibinfo{author}{\bibfnamefont{D.~L.} \bibnamefont{Turcotte}},
  \emph{\bibinfo{title}{Fractals and Chaos in Geology and Geophysics}}
  (\bibinfo{publisher}{Cambridge University Press},
  \bibinfo{address}{Cambridge}, \bibinfo{year}{1997}), \bibinfo{edition}{2nd}
  ed.

\bibitem[{\citenamefont{Holliday et~al.}(2006)\citenamefont{Holliday, Rundle,
  Turcotte, Klein, Tiampo, and Donnellan}}]{HOL06}
\bibinfo{author}{\bibfnamefont{J.~R.} \bibnamefont{Holliday}},
  \bibinfo{author}{\bibfnamefont{J.~B.} \bibnamefont{Rundle}},
  \bibinfo{author}{\bibfnamefont{D.~L.} \bibnamefont{Turcotte}},
  \bibinfo{author}{\bibfnamefont{W.}~\bibnamefont{Klein}},
  \bibinfo{author}{\bibfnamefont{K.~F.} \bibnamefont{Tiampo}},
  \bibnamefont{and}
  \bibinfo{author}{\bibfnamefont{A.}~\bibnamefont{Donnellan}},
  \bibinfo{journal}{Phys. Rev. Lett.} \textbf{\bibinfo{volume}{97}},
  \bibinfo{pages}{238501} (\bibinfo{year}{2006}).

\bibitem[{\citenamefont{Varotsos et~al.}(2005)\citenamefont{Varotsos, Sarlis,
  Tanaka, and Skordas}}]{NAT05C}
\bibinfo{author}{\bibfnamefont{P.~A.} \bibnamefont{Varotsos}},
  \bibinfo{author}{\bibfnamefont{N.~V.} \bibnamefont{Sarlis}},
  \bibinfo{author}{\bibfnamefont{H.~K.} \bibnamefont{Tanaka}},
  \bibnamefont{and} \bibinfo{author}{\bibfnamefont{E.~S.}
  \bibnamefont{Skordas}}, \bibinfo{journal}{Phys. Rev. E}
  \textbf{\bibinfo{volume}{72}}, \bibinfo{pages}{041103}
  (\bibinfo{year}{2005}).

\bibitem[{\citenamefont{Varotsos et~al.}(2013)\citenamefont{Varotsos, Sarlis,
  Skordas, and Lazaridou}}]{TECTO12}
\bibinfo{author}{\bibfnamefont{P.~A.} \bibnamefont{Varotsos}},
  \bibinfo{author}{\bibfnamefont{N.~V.} \bibnamefont{Sarlis}},
  \bibinfo{author}{\bibfnamefont{E.~S.} \bibnamefont{Skordas}},
  \bibnamefont{and} \bibinfo{author}{\bibfnamefont{M.~S.}
  \bibnamefont{Lazaridou}}, \bibinfo{journal}{Tectonophysics}
  \textbf{\bibinfo{volume}{589}}, \bibinfo{pages}{116} (\bibinfo{year}{2013}).

\bibitem[{\citenamefont{{Japan Meteorological Agency}}(2000)}]{JMA00}
\bibinfo{author}{\bibnamefont{{Japan Meteorological Agency}}},
  \bibinfo{journal}{Earth Planets and Space} \textbf{\bibinfo{volume}{52}},
  \bibinfo{pages}{i} (\bibinfo{year}{2000}).

\bibitem[{\citenamefont{Sarlis et~al.}(2013)\citenamefont{Sarlis, Skordas,
  Varotsos, Nagao, Kamogawa, Tanaka, and Uyeda}}]{PNAS13}
\bibinfo{author}{\bibfnamefont{N.~V.} \bibnamefont{Sarlis}},
  \bibinfo{author}{\bibfnamefont{E.~S.} \bibnamefont{Skordas}},
  \bibinfo{author}{\bibfnamefont{P.~A.} \bibnamefont{Varotsos}},
  \bibinfo{author}{\bibfnamefont{T.}~\bibnamefont{Nagao}},
  \bibinfo{author}{\bibfnamefont{M.}~\bibnamefont{Kamogawa}},
  \bibinfo{author}{\bibfnamefont{H.}~\bibnamefont{Tanaka}}, \bibnamefont{and}
  \bibinfo{author}{\bibfnamefont{S.}~\bibnamefont{Uyeda}},
  \bibinfo{journal}{Proc. Natl. Acad. Sci. USA} \textbf{\bibinfo{volume}{110}},
  \bibinfo{pages}{13734} (\bibinfo{year}{2013}).

\bibitem[{\citenamefont{Yamanaka and Kikuchi}(2004)}]{YAM04}
\bibinfo{author}{\bibfnamefont{Y.}~\bibnamefont{Yamanaka}} \bibnamefont{and}
  \bibinfo{author}{\bibfnamefont{M.}~\bibnamefont{Kikuchi}},
  \bibinfo{journal}{Journal of Geophysical Research: Solid Earth}
  \textbf{\bibinfo{volume}{109}}, \bibinfo{pages}{B07307}
  (\bibinfo{year}{2004}), ISSN \bibinfo{issn}{2156-2202}.

\bibitem[{\citenamefont{Peng et~al.}(1994)\citenamefont{Peng, Buldyrev, Havlin,
  Simons, Stanley, and Goldberger}}]{PEN94}
\bibinfo{author}{\bibfnamefont{C.-K.} \bibnamefont{Peng}},
  \bibinfo{author}{\bibfnamefont{S.~V.} \bibnamefont{Buldyrev}},
  \bibinfo{author}{\bibfnamefont{S.}~\bibnamefont{Havlin}},
  \bibinfo{author}{\bibfnamefont{M.}~\bibnamefont{Simons}},
  \bibinfo{author}{\bibfnamefont{H.~E.} \bibnamefont{Stanley}},
  \bibnamefont{and} \bibinfo{author}{\bibfnamefont{A.~L.}
  \bibnamefont{Goldberger}}, \bibinfo{journal}{Phys. Rev. E}
  \textbf{\bibinfo{volume}{49}}, \bibinfo{pages}{1685} (\bibinfo{year}{1994}).

\bibitem[{\citenamefont{Peng et~al.}(1993)\citenamefont{Peng, Buldyrev,
  Goldberger, Havlin, Simons, and Stanley}}]{PEN93}
\bibinfo{author}{\bibfnamefont{C.-K.} \bibnamefont{Peng}},
  \bibinfo{author}{\bibfnamefont{S.~V.} \bibnamefont{Buldyrev}},
  \bibinfo{author}{\bibfnamefont{A.~L.} \bibnamefont{Goldberger}},
  \bibinfo{author}{\bibfnamefont{S.}~\bibnamefont{Havlin}},
  \bibinfo{author}{\bibfnamefont{M.}~\bibnamefont{Simons}}, \bibnamefont{and}
  \bibinfo{author}{\bibfnamefont{H.~E.} \bibnamefont{Stanley}},
  \bibinfo{journal}{Phys. Rev. E} \textbf{\bibinfo{volume}{47}},
  \bibinfo{pages}{3730} (\bibinfo{year}{1993}).

\bibitem[{\citenamefont{Peng et~al.}(1995)\citenamefont{Peng, Havlin, Stanley,
  and Goldberger}}]{PEN95B}
\bibinfo{author}{\bibfnamefont{C.~K.} \bibnamefont{Peng}},
  \bibinfo{author}{\bibfnamefont{S.}~\bibnamefont{Havlin}},
  \bibinfo{author}{\bibfnamefont{H.~E.} \bibnamefont{Stanley}},
  \bibnamefont{and} \bibinfo{author}{\bibfnamefont{A.~L.}
  \bibnamefont{Goldberger}}, \bibinfo{journal}{CHAOS}
  \textbf{\bibinfo{volume}{5}}, \bibinfo{pages}{82} (\bibinfo{year}{1995}).

\bibitem[{\citenamefont{Ashkenazy et~al.}(2002)\citenamefont{Ashkenazy,
  Hausdorff, Ivanov, and Stanley}}]{ASH02}
\bibinfo{author}{\bibfnamefont{Y.}~\bibnamefont{Ashkenazy}},
  \bibinfo{author}{\bibfnamefont{J.~M.} \bibnamefont{Hausdorff}},
  \bibinfo{author}{\bibfnamefont{P.~C.} \bibnamefont{Ivanov}},
  \bibnamefont{and} \bibinfo{author}{\bibfnamefont{H.~E.}
  \bibnamefont{Stanley}}, \bibinfo{journal}{Physica A}
  \textbf{\bibinfo{volume}{316}}, \bibinfo{pages}{662} (\bibinfo{year}{2002}).

\bibitem[{\citenamefont{Ivanov et~al.}(2009)\citenamefont{Ivanov, Ma, Bartsch,
  Hausdorff, Nunes~Amaral, Schulte-Frohlinde, Stanley, and Yoneyama}}]{IVA09}
\bibinfo{author}{\bibfnamefont{P.~C.} \bibnamefont{Ivanov}},
  \bibinfo{author}{\bibfnamefont{Q.~D.~Y.} \bibnamefont{Ma}},
  \bibinfo{author}{\bibfnamefont{R.~P.} \bibnamefont{Bartsch}},
  \bibinfo{author}{\bibfnamefont{J.~M.} \bibnamefont{Hausdorff}},
  \bibinfo{author}{\bibfnamefont{L.~A.} \bibnamefont{Nunes~Amaral}},
  \bibinfo{author}{\bibfnamefont{V.}~\bibnamefont{Schulte-Frohlinde}},
  \bibinfo{author}{\bibfnamefont{H.~E.} \bibnamefont{Stanley}},
  \bibnamefont{and} \bibinfo{author}{\bibfnamefont{M.}~\bibnamefont{Yoneyama}},
  \bibinfo{journal}{Phys. Rev. E} \textbf{\bibinfo{volume}{79}},
  \bibinfo{pages}{041920} (\bibinfo{year}{2009}).

\bibitem[{\citenamefont{Ivanov}(2007)}]{IVA07B}
\bibinfo{author}{\bibfnamefont{P.~C.} \bibnamefont{Ivanov}},
  \bibinfo{journal}{IEEE Eng. Med. Biol.} \textbf{\bibinfo{volume}{26}},
  \bibinfo{pages}{33} (\bibinfo{year}{2007}).

\bibitem[{\citenamefont{Talkner and Weber}(2000)}]{TAL00}
\bibinfo{author}{\bibfnamefont{P.}~\bibnamefont{Talkner}} \bibnamefont{and}
  \bibinfo{author}{\bibfnamefont{R.~O.} \bibnamefont{Weber}},
  \bibinfo{journal}{Phys. Rev. E} \textbf{\bibinfo{volume}{62}},
  \bibinfo{pages}{150} (\bibinfo{year}{2000}).

\bibitem[{\citenamefont{Goldberger et~al.}(2002)\citenamefont{Goldberger,
  Amaral, Hausdorff, Ivanov, Peng, and Stanley}}]{GOL02}
\bibinfo{author}{\bibfnamefont{A.~L.} \bibnamefont{Goldberger}},
  \bibinfo{author}{\bibfnamefont{L.~A.~N.} \bibnamefont{Amaral}},
  \bibinfo{author}{\bibfnamefont{J.~M.} \bibnamefont{Hausdorff}},
  \bibinfo{author}{\bibfnamefont{P.~C.} \bibnamefont{Ivanov}},
  \bibinfo{author}{\bibfnamefont{C.-K.} \bibnamefont{Peng}}, \bibnamefont{and}
  \bibinfo{author}{\bibfnamefont{H.~E.} \bibnamefont{Stanley}},
  \bibinfo{journal}{Proc. Natl. Acad. Sci. USA} \textbf{\bibinfo{volume}{99}},
  \bibinfo{pages}{2466} (\bibinfo{year}{2002}).

\bibitem[{\citenamefont{Telesca and Lovallo}(2009)}]{TEL09}
\bibinfo{author}{\bibfnamefont{L.}~\bibnamefont{Telesca}} \bibnamefont{and}
  \bibinfo{author}{\bibfnamefont{M.}~\bibnamefont{Lovallo}},
  \bibinfo{journal}{Geophys. Res. Lett.} \textbf{\bibinfo{volume}{36}},
  \bibinfo{pages}{L01308} (\bibinfo{year}{2009}), ISSN
  \bibinfo{issn}{1944-8007}.

\bibitem[{\citenamefont{Telesca and Lasaponara}(2006)}]{TEL06}
\bibinfo{author}{\bibfnamefont{L.}~\bibnamefont{Telesca}} \bibnamefont{and}
  \bibinfo{author}{\bibfnamefont{R.}~\bibnamefont{Lasaponara}},
  \bibinfo{journal}{Geophysical Research Letters}
  \textbf{\bibinfo{volume}{33}}, \bibinfo{pages}{L14401}
  (\bibinfo{year}{2006}), ISSN \bibinfo{issn}{1944-8007}.

\bibitem[{\citenamefont{Telesca et~al.}(2012)\citenamefont{Telesca, Pierini,
  and Scian}}]{TEL12}
\bibinfo{author}{\bibfnamefont{L.}~\bibnamefont{Telesca}},
  \bibinfo{author}{\bibfnamefont{J.~O.} \bibnamefont{Pierini}},
  \bibnamefont{and} \bibinfo{author}{\bibfnamefont{B.}~\bibnamefont{Scian}},
  \bibinfo{journal}{Physica A: Statistical Mechanics and its Applications}
  \textbf{\bibinfo{volume}{391}}, \bibinfo{pages}{1553 }
  (\bibinfo{year}{2012}), ISSN \bibinfo{issn}{0378-4371}.

\bibitem[{\citenamefont{Rong et~al.}(2012)\citenamefont{Rong, Wang, Ding, and
  Huang}}]{RON12}
\bibinfo{author}{\bibfnamefont{Y.}~\bibnamefont{Rong}},
  \bibinfo{author}{\bibfnamefont{Q.}~\bibnamefont{Wang}},
  \bibinfo{author}{\bibfnamefont{X.}~\bibnamefont{Ding}}, \bibnamefont{and}
  \bibinfo{author}{\bibfnamefont{Q.}~\bibnamefont{Huang}},
  \bibinfo{journal}{Chinese Journal of Geophysics}
  \textbf{\bibinfo{volume}{55}}, \bibinfo{eid}{3709}
  (pages~\bibinfo{numpages}{8}) (\bibinfo{year}{2012}).

\bibitem[{\citenamefont{Shao et~al.}(2012)\citenamefont{Shao, Gu, Zhou, and
  Sornette}}]{SHAO12}
\bibinfo{author}{\bibfnamefont{Y.-H.} \bibnamefont{Shao}},
  \bibinfo{author}{\bibfnamefont{G.-F.} \bibnamefont{Gu}},
  \bibinfo{author}{\bibfnamefont{W.-X.} \bibnamefont{Zhou}}, \bibnamefont{and}
  \bibinfo{author}{\bibfnamefont{D.}~\bibnamefont{Sornette}},
  \bibinfo{journal}{Sci. Rep.} \textbf{\bibinfo{volume}{2}},
  \bibinfo{pages}{835} (\bibinfo{year}{2012}).

\bibitem[{\citenamefont{Varotsos et~al.}(2001)\citenamefont{Varotsos, Sarlis,
  and Skordas}}]{NAT01}
\bibinfo{author}{\bibfnamefont{P.~A.} \bibnamefont{Varotsos}},
  \bibinfo{author}{\bibfnamefont{N.~V.} \bibnamefont{Sarlis}},
  \bibnamefont{and} \bibinfo{author}{\bibfnamefont{E.~S.}
  \bibnamefont{Skordas}}, \bibinfo{journal}{Practica of Athens Academy}
  \textbf{\bibinfo{volume}{76}}, \bibinfo{pages}{294} (\bibinfo{year}{2001}).

\bibitem[{\citenamefont{Varotsos
  et~al.}(2002{\natexlab{a}})\citenamefont{Varotsos, Sarlis, and
  Skordas}}]{NAT02}
\bibinfo{author}{\bibfnamefont{P.~A.} \bibnamefont{Varotsos}},
  \bibinfo{author}{\bibfnamefont{N.~V.} \bibnamefont{Sarlis}},
  \bibnamefont{and} \bibinfo{author}{\bibfnamefont{E.~S.}
  \bibnamefont{Skordas}}, \bibinfo{journal}{Phys. Rev. E}
  \textbf{\bibinfo{volume}{66}}, \bibinfo{pages}{011902}
  (\bibinfo{year}{2002}{\natexlab{a}}).

\bibitem[{\citenamefont{Varotsos
  et~al.}(2002{\natexlab{b}})\citenamefont{Varotsos, Sarlis, and
  Skordas}}]{NAT02A}
\bibinfo{author}{\bibfnamefont{P.~A.} \bibnamefont{Varotsos}},
  \bibinfo{author}{\bibfnamefont{N.~V.} \bibnamefont{Sarlis}},
  \bibnamefont{and} \bibinfo{author}{\bibfnamefont{E.~S.}
  \bibnamefont{Skordas}}, \bibinfo{journal}{Acta Geophysica Polonica}
  \textbf{\bibinfo{volume}{50}}, \bibinfo{pages}{337}
  (\bibinfo{year}{2002}{\natexlab{b}}).

\bibitem[{\citenamefont{Abe et~al.}(2005)\citenamefont{Abe, Sarlis, Skordas,
  Tanaka, and Varotsos}}]{ABE05}
\bibinfo{author}{\bibfnamefont{S.}~\bibnamefont{Abe}},
  \bibinfo{author}{\bibfnamefont{N.~V.} \bibnamefont{Sarlis}},
  \bibinfo{author}{\bibfnamefont{E.~S.} \bibnamefont{Skordas}},
  \bibinfo{author}{\bibfnamefont{H.~K.} \bibnamefont{Tanaka}},
  \bibnamefont{and} \bibinfo{author}{\bibfnamefont{P.~A.}
  \bibnamefont{Varotsos}}, \bibinfo{journal}{Phys. Rev. Lett.}
  \textbf{\bibinfo{volume}{94}}, \bibinfo{pages}{170601}
  (\bibinfo{year}{2005}).

\bibitem[{\citenamefont{Varotsos
  et~al.}(2011{\natexlab{b}})\citenamefont{Varotsos, Sarlis, Skordas, Uyeda,
  and Kamogawa}}]{PNAS}
\bibinfo{author}{\bibfnamefont{P.}~\bibnamefont{Varotsos}},
  \bibinfo{author}{\bibfnamefont{N.~V.} \bibnamefont{Sarlis}},
  \bibinfo{author}{\bibfnamefont{E.~S.} \bibnamefont{Skordas}},
  \bibinfo{author}{\bibfnamefont{S.}~\bibnamefont{Uyeda}}, \bibnamefont{and}
  \bibinfo{author}{\bibfnamefont{M.}~\bibnamefont{Kamogawa}},
  \bibinfo{journal}{Proc. Natl. Acad. Sci. USA} \textbf{\bibinfo{volume}{108}},
  \bibinfo{pages}{11361} (\bibinfo{year}{2011}{\natexlab{b}}).

\bibitem[{\citenamefont{Varotsos
  et~al.}(2011{\natexlab{c}})\citenamefont{Varotsos, Sarlis, and
  Skordas}}]{VAR11}
\bibinfo{author}{\bibfnamefont{P.}~\bibnamefont{Varotsos}},
  \bibinfo{author}{\bibfnamefont{N.}~\bibnamefont{Sarlis}}, \bibnamefont{and}
  \bibinfo{author}{\bibfnamefont{E.}~\bibnamefont{Skordas}},
  \bibinfo{journal}{EPL} \textbf{\bibinfo{volume}{96}}, \bibinfo{pages}{59002}
  (\bibinfo{year}{2011}{\natexlab{c}}).

\bibitem[{\citenamefont{Sarlis et~al.}(2010{\natexlab{a}})\citenamefont{Sarlis,
  Skordas, and Varotsos}}]{NEWEPL}
\bibinfo{author}{\bibfnamefont{N.~V.} \bibnamefont{Sarlis}},
  \bibinfo{author}{\bibfnamefont{E.~S.} \bibnamefont{Skordas}},
  \bibnamefont{and} \bibinfo{author}{\bibfnamefont{P.~A.}
  \bibnamefont{Varotsos}}, \bibinfo{journal}{EPL}
  \textbf{\bibinfo{volume}{91}}, \bibinfo{pages}{59001}
  (\bibinfo{year}{2010}{\natexlab{a}}).

\bibitem[{\citenamefont{Goldberger et~al.}(2000)\citenamefont{Goldberger,
  Amaral, Glass, Hausdorff, Ivanov, Mark, Mictus, Moody, Peng, and
  Stanley}}]{GOL00}
\bibinfo{author}{\bibfnamefont{A.~L.} \bibnamefont{Goldberger}},
  \bibinfo{author}{\bibfnamefont{L.~A.~N.} \bibnamefont{Amaral}},
  \bibinfo{author}{\bibfnamefont{L.}~\bibnamefont{Glass}},
  \bibinfo{author}{\bibfnamefont{J.~M.} \bibnamefont{Hausdorff}},
  \bibinfo{author}{\bibfnamefont{P.~C.} \bibnamefont{Ivanov}},
  \bibinfo{author}{\bibfnamefont{R.~G.} \bibnamefont{Mark}},
  \bibinfo{author}{\bibfnamefont{J.~E.} \bibnamefont{Mictus}},
  \bibinfo{author}{\bibfnamefont{G.~B.} \bibnamefont{Moody}},
  \bibinfo{author}{\bibfnamefont{C.-K.} \bibnamefont{Peng}}, \bibnamefont{and}
  \bibinfo{author}{\bibfnamefont{H.~E.} \bibnamefont{Stanley}},
  \bibinfo{journal}{Circulation} \textbf{\bibinfo{volume}{101}},
  \bibinfo{pages}{E215 (see also {\tt www.physionet.org})}
  (\bibinfo{year}{2000}).

\bibitem[{\citenamefont{Tenenbaum et~al.}(2012)\citenamefont{Tenenbaum, Havlin,
  and Stanley}}]{TEN12}
\bibinfo{author}{\bibfnamefont{J.~N.} \bibnamefont{Tenenbaum}},
  \bibinfo{author}{\bibfnamefont{S.}~\bibnamefont{Havlin}}, \bibnamefont{and}
  \bibinfo{author}{\bibfnamefont{H.~E.} \bibnamefont{Stanley}},
  \bibinfo{journal}{Phys. Rev. E} \textbf{\bibinfo{volume}{86}},
  \bibinfo{pages}{046107} (\bibinfo{year}{2012}).

\bibitem[{\citenamefont{Tanaka et~al.}(2004)\citenamefont{Tanaka, Varotsos,
  Sarlis, and Skordas}}]{TAN04}
\bibinfo{author}{\bibfnamefont{H.~K.} \bibnamefont{Tanaka}},
  \bibinfo{author}{\bibfnamefont{P.~V.} \bibnamefont{Varotsos}},
  \bibinfo{author}{\bibfnamefont{N.~V.} \bibnamefont{Sarlis}},
  \bibnamefont{and} \bibinfo{author}{\bibfnamefont{E.~S.}
  \bibnamefont{Skordas}}, \bibinfo{journal}{Proc. Japan Acad., Ser. B}
  \textbf{\bibinfo{volume}{80}}, \bibinfo{pages}{283} (\bibinfo{year}{2004}).

\bibitem[{\citenamefont{Kanamori}(1978)}]{KAN78}
\bibinfo{author}{\bibfnamefont{H.}~\bibnamefont{Kanamori}},
  \bibinfo{journal}{Nature} \textbf{\bibinfo{volume}{271}},
  \bibinfo{pages}{411} (\bibinfo{year}{1978}).

\bibitem[{\citenamefont{Xu et~al.}(2013)\citenamefont{Xu, Han, Huang, Hattori,
  Febriani, and Yamaguchi}}]{XU013}
\bibinfo{author}{\bibfnamefont{G.}~\bibnamefont{Xu}},
  \bibinfo{author}{\bibfnamefont{P.}~\bibnamefont{Han}},
  \bibinfo{author}{\bibfnamefont{Q.}~\bibnamefont{Huang}},
  \bibinfo{author}{\bibfnamefont{K.}~\bibnamefont{Hattori}},
  \bibinfo{author}{\bibfnamefont{F.}~\bibnamefont{Febriani}}, \bibnamefont{and}
  \bibinfo{author}{\bibfnamefont{H.}~\bibnamefont{Yamaguchi}},
  \bibinfo{journal}{J. Asian Earth Sci.} \textbf{\bibinfo{volume}{77}},
  \bibinfo{pages}{59 } (\bibinfo{year}{2013}), ISSN \bibinfo{issn}{1367-9120},
  \urlprefix\url{http://www.sciencedirect.com/science/article/pii/S13679120130%
04197}.

\bibitem[{\citenamefont{Varotsos
  et~al.}(2003{\natexlab{c}})\citenamefont{Varotsos, Sarlis, and
  Skordas}}]{VAR03}
\bibinfo{author}{\bibfnamefont{P.~V.} \bibnamefont{Varotsos}},
  \bibinfo{author}{\bibfnamefont{N.~V.} \bibnamefont{Sarlis}},
  \bibnamefont{and} \bibinfo{author}{\bibfnamefont{E.~S.}
  \bibnamefont{Skordas}}, \bibinfo{journal}{Phys. Rev. Lett.}
  \textbf{\bibinfo{volume}{91}}, \bibinfo{pages}{148501}
  (\bibinfo{year}{2003}{\natexlab{c}}).

\bibitem[{\citenamefont{Telesca et~al.}(2003)\citenamefont{Telesca, Lapenna,
  and Macchiato}}]{TEL03}
\bibinfo{author}{\bibfnamefont{L.}~\bibnamefont{Telesca}},
  \bibinfo{author}{\bibfnamefont{V.}~\bibnamefont{Lapenna}}, \bibnamefont{and}
  \bibinfo{author}{\bibfnamefont{M.}~\bibnamefont{Macchiato}},
  \bibinfo{journal}{Earth and Planetary Science Letters}
  \textbf{\bibinfo{volume}{212}}, \bibinfo{pages}{279 } (\bibinfo{year}{2003}),
  ISSN \bibinfo{issn}{0012-821X}.

\bibitem[{\citenamefont{Lennartz et~al.}(2008)\citenamefont{Lennartz, Livina,
  Bunde, and Havlin}}]{LEN08}
\bibinfo{author}{\bibfnamefont{S.}~\bibnamefont{Lennartz}},
  \bibinfo{author}{\bibfnamefont{V.~N.} \bibnamefont{Livina}},
  \bibinfo{author}{\bibfnamefont{A.}~\bibnamefont{Bunde}}, \bibnamefont{and}
  \bibinfo{author}{\bibfnamefont{S.}~\bibnamefont{Havlin}},
  \bibinfo{journal}{EPL} \textbf{\bibinfo{volume}{81}}, \bibinfo{pages}{69001}
  (\bibinfo{year}{2008}).

\bibitem[{\citenamefont{Sarlis et~al.}(2010{\natexlab{b}})\citenamefont{Sarlis,
  Skordas, and Varotsos}}]{NEWTSA}
\bibinfo{author}{\bibfnamefont{N.~V.} \bibnamefont{Sarlis}},
  \bibinfo{author}{\bibfnamefont{E.~S.} \bibnamefont{Skordas}},
  \bibnamefont{and} \bibinfo{author}{\bibfnamefont{P.~A.}
  \bibnamefont{Varotsos}}, \bibinfo{journal}{Phys. Rev. E}
  \textbf{\bibinfo{volume}{82}}, \bibinfo{pages}{021110}
  (\bibinfo{year}{2010}{\natexlab{b}}).

\bibitem[{\citenamefont{Abe and Okamoto}(2001)}]{ABE01}
\bibinfo{editor}{\bibfnamefont{S.}~\bibnamefont{Abe}} \bibnamefont{and}
  \bibinfo{editor}{\bibfnamefont{Y.}~\bibnamefont{Okamoto}}, eds.,
  \emph{\bibinfo{title}{Non Extensive Statistical Mechanics and its
  Applications}} (\bibinfo{publisher}{Springer}, \bibinfo{address}{Berlin},
  \bibinfo{year}{2001}).

\bibitem[{\citenamefont{Tsallis}(2009)}]{TSA09}
\bibinfo{author}{\bibfnamefont{C.}~\bibnamefont{Tsallis}},
  \emph{\bibinfo{title}{Introduction to Nonextensive Statistical Mechanics}}
  (\bibinfo{publisher}{Springer}, \bibinfo{address}{Berlin},
  \bibinfo{year}{2009}).

\bibitem[{\citenamefont{Tsallis}(1988)}]{TSA88}
\bibinfo{author}{\bibfnamefont{C.}~\bibnamefont{Tsallis}}, \bibinfo{journal}{J.
  Stat. Phys.} \textbf{\bibinfo{volume}{52}}, \bibinfo{pages}{479}
  (\bibinfo{year}{1988}).

\bibitem[{\citenamefont{Livadiotis and McComas}(2009)}]{LIV09}
\bibinfo{author}{\bibfnamefont{G.}~\bibnamefont{Livadiotis}} \bibnamefont{and}
  \bibinfo{author}{\bibfnamefont{D.~J.} \bibnamefont{McComas}},
  \bibinfo{journal}{Journal of Geophysical Research: Space Physics}
  \textbf{\bibinfo{volume}{114}}, \bibinfo{pages}{A11105}
  (\bibinfo{year}{2009}), ISSN \bibinfo{issn}{2156-2202},
  \urlprefix\url{10.1029/2009JA014352}.

\bibitem[{\citenamefont{Telesca}(2012)}]{TEL12A}
\bibinfo{author}{\bibfnamefont{L.}~\bibnamefont{Telesca}},
  \bibinfo{journal}{Bulletin of the Seismological Society of America}
  \textbf{\bibinfo{volume}{102}}, \bibinfo{pages}{886} (\bibinfo{year}{2012}),
  \eprint{http://www.bssaonline.org/content/102/2/886.full.pdf+html}.

\bibitem[{\citenamefont{Varotsos et~al.}(2012)\citenamefont{Varotsos, Sarlis,
  and Skordas}}]{EPL12}
\bibinfo{author}{\bibfnamefont{P.}~\bibnamefont{Varotsos}},
  \bibinfo{author}{\bibfnamefont{N.}~\bibnamefont{Sarlis}}, \bibnamefont{and}
  \bibinfo{author}{\bibfnamefont{E.}~\bibnamefont{Skordas}},
  \bibinfo{journal}{EPL} \textbf{\bibinfo{volume}{99}}, \bibinfo{pages}{59001}
  (\bibinfo{year}{2012}).

\bibitem[{\citenamefont{Varotsos et~al.}(2007)\citenamefont{Varotsos, Sarlis,
  Skordas, and Lazaridou}}]{NAT07}
\bibinfo{author}{\bibfnamefont{P.~A.} \bibnamefont{Varotsos}},
  \bibinfo{author}{\bibfnamefont{N.~V.} \bibnamefont{Sarlis}},
  \bibinfo{author}{\bibfnamefont{E.~S.} \bibnamefont{Skordas}},
  \bibnamefont{and} \bibinfo{author}{\bibfnamefont{M.~S.}
  \bibnamefont{Lazaridou}}, \bibinfo{journal}{Appl. Phys. Lett.}
  \textbf{\bibinfo{volume}{91}}, \bibinfo{eid}{064106} (\bibinfo{year}{2007}).

\end{thebibliography}
\end{document}